\documentclass[twocolumn,tighten]{aastex63}
\hypersetup{linkcolor=red,citecolor=blue,filecolor=cyan,urlcolor=magenta}
\usepackage{amsmath}
\usepackage{natbib}
\usepackage{CJK}

\newcommand{\degree}{$^{\circ}$}
\newcommand{\loggf}{\mbox{$\log gf$}}
\newcommand{\kmsec}{\mbox{km~s$^{\rm -1}$}}
\newcommand{\logg}{\mbox{log~{\it g}}}

\newcommand{\teff}{\mbox{$T_{\rm eff}$}}
\newcommand{\vt}{\mbox{$v_{\rm t}$}}

\shorttitle{Vanadium in Metal-Poor Stars}
\shortauthors{Ou et al.}
\accepted{for publication in the Astrophysical Journal}

\begin{document}
\begin{CJK*}{UTF8}{gbsn}

\title{%
Vanadium Abundance Derivations in 255 Metal-poor Stars\footnote{%
This paper includes data gathered with the 6.5 meter
Magellan Telescopes located at Las Campanas Observatory, Chile,
and The McDonald Observatory of The University of Texas at Austin.
The Hobby-Eberly Telescope is a
joint project of the University of Texas at Austin, the Pennsylvania State
University, Stanford University, Ludwig-Maximilians-Universit\"{a}t
M\"{u}nchen, and Georg-August-Universit\"{a}t G\"{o}ttingen.}
}

\author{Xiaowei Ou (欧筱葳)}
\affiliation{%
Department of Astronomy, University of Michigan,
1085 S.\ University Ave., Ann Arbor, MI 48109, USA}
\email{Email:\ ouxw@umich.edu}

\author{Ian U.\ Roederer}
\affiliation{%
Department of Astronomy, University of Michigan,
1085 S.\ University Ave., Ann Arbor, MI 48109, USA}
\affiliation{%
Joint Institute for Nuclear Astrophysics -- Center for the
Evolution of the Elements (JINA-CEE), USA}

\author{Christopher Sneden}
\affiliation{%
Department of Astronomy and McDonald Observatory, University of Texas, 
2515 Speedway, Stop C1400, Austin, TX 78712, USA}

\author{John J.\ Cowan}
\affiliation{%
HLD Department of Physics \& Astronomy, University of Oklahoma,
440 W. Brooks St., Norman, OK 73019, USA}

\author{James E.\ Lawler}
\affiliation{%
Physics Department, University of Wisconsin-Madison,
1150 University Avenue, Madison, WI 53706-1390, USA}

\author{Stephen A. Shectman}
\affiliation{%
Carnegie Observatories, 
813 Santa Barbara Street, Pasadena, CA 91101, USA}

\author{Ian B. Thompson}
\affiliation{%
Carnegie Observatories, 
813 Santa Barbara Street, Pasadena, CA 91101, USA}

\begin{abstract}

We present
vanadium (V) abundances for 255
metal-poor stars,
derived from 
high-resolution optical spectra
from the Magellan Inamori Kyocera Echelle spectrograph on the Magellan Telescopes at Las Campanas
Observatory, the Robert G.\ Tull Coud\'{e}
Spectrograph on the Harlan J.\ Smith Telescope at McDonald Observatory, and
the High Resolution Spectrograph on the Hobby-Eberly Telescope at McDonald Observatory.
We use 
updated V~\textsc{i} and V~\textsc{ii} atomic transition
data from 
recent laboratory studies, 
and we increase the number of lines examined 
(from 1 to 4 lines of V~\textsc{i}, and 
from 2 to 7 lines of V~\textsc{ii}).
As a result, we reduce the V abundance uncertainties 
for most stars by more than 20~\%
and expand the number of stars with V detections from 204 to 255.
In the metallicity range $-$4.0~$<$~[Fe/H]~$< -$1.0,
we calculate the mean ratios
[V~\textsc{i}/Fe~\textsc{i}]~$= -0.10 \pm 0.01~(\sigma = 0.16)$
from 128 stars with $\geq$~2 V~\textsc{i} lines detected,
[V~\textsc{ii}/Fe~\textsc{ii}]~$= +0.13 \pm 0.01~(\sigma = 0.16)$
from 220 stars with $\geq$~2 V~\textsc{ii} lines detected, 
and
[V~\textsc{ii}/V~\textsc{i}]~$= +0.25 \pm 0.01~(\sigma = 0.15)$
from 119 stars.
We suspect this offset is due to non-LTE effects, and we
recommend using [V~\textsc{ii}/Fe~\textsc{ii}], which
is enhanced relative to the solar ratio, as a better representation
of [V/Fe].
We provide more extensive evidence for abundance correlations detected previously among scandium, titanium, and vanadium, and we
identify no systematic effects in the analysis that can
explain these correlations.

\end{abstract}



\section{Introduction}
\label{intro}
\end{CJK*}

Metal-poor stars in the halo record the early chemical enrichment history of the Galaxy.
With only a few prior stellar explosions pouring metals into the environment,
their chemical compositions tell us about the nucleosynthetic processes responsible
for creating many of the heavy elements in the universe.
Many theoretical supernova (SN) models have been used to provide predictions about the
relative chemical abundances in these metal-poor stars.

In particular, the nucleosynthesis of iron-group elements has
been studied in detail to inform Galactic chemical evolution models (e.g., \citealt{kobayashi11,curtis19}).
In general, these elements are produced mostly
in core-collapse supernovae (CCSNe) and explosive 
stellar yields.
Specifically, scandium (Sc, $Z = 21$), titanium (Ti, $Z = 22$), and vanadium (V, $Z = 23$)
are produced by explosive silicon burning and oxygen burning 
in the core collapse SN phase,
while the other iron-group elements are also 
produced by incomplete or complete silicon burning in the 
CCSNe ejecta \citep{woosley95}.
While significant developments in abundance derivations have been made over the past decade,
challenges exist for obtaining reliable abundances for these elements.
Vanadium, along with other iron-group elements 
(21 $\leq Z \leq$ 28), is present in the atmospheres of metal-poor
stars mostly in the ionized state. 
These ionized-species transitions usually require 
blue or ultraviolet (UV) spectra for detection.
They also require 
accurate laboratory atomic transition data to yield reliable abundances.
Using accurate and precise transition probabilities and hyperfine/isotopic substructure atomic data from the 
Wisconsin atomic physics group, \citet{sneden16} derived iron-group 
element abundances for the very metal-poor star HD~84937.
They discovered that the abundances of scandium, titanium, and vanadium were all enhanced, relative to iron (Fe, $Z = 26$) in HD~84937, 
and they confirmed that similar correlations exist among larger samples of metal-poor stars using data 
from \citet{cayrel04}, \citet{cohen04,cohen08emp}, \citet{barklem05heres}, \citet{lai08}, \citet{yong13a}, and \citet{roederer14c}.
Recently, \citet{cowan20} reported iron-group abundances for three
main-sequence turnoff stars with metallicities [Fe/H] $\sim -3$, confirming and
extending the trends found in \citet{sneden16}.

So far, current CCSN models cannot fully explain
the correlations observed in the halo metal poor stars.
\citet{sneden16} described 
the possible common production process for 
titanium and vanadium, but they raised questions
on the theoretical reasons for the correlation between scandium
and the other two elements. 
Not only is it unclear how the complex production history of scandium
ties to its correlation with vanadium and titanium,
but the abundance of scandium itself is also difficult to reproduce theoretically.
Our current Galactic chemical evolution (GCE) models,
in general, underestimate the abundances of scandium, titanium, and
vanadium in metal-poor stars \citep{kobayashi06}. 

It has been suggested
that a 50~\%\ hypernova fraction at early Galactic times may be able
to explain the overall overabundance of these elements
\citep{umeda02,kobayashi06}.
\citeauthor{umeda02} calculated nucleosynthesis with CCSNe 
models and compare them with observations to constrain the parameters,
with an emphasis on trying to recover the high iron-group element production.
They find that deep mixing, deep mass cuts, smaller neutron excesses, 
and larger explosion energies produce higher iron-group abundances in general.
Yet, even then scandium is still underproduced, 
although a better match could potentially be obtained
by introducing jets and neutrino effects
\citep{kobayashi06,kobayashi11}.
Following that direction,
\citet{curtis19} managed to reproduce most of the abundance
measurements of the very metal-poor star HD~84937, studied in
\citet{sneden16}, using CCSNe models that consistently capture
the neutrino interactions, but with large variations in scandium
and zinc.
It is also unclear how these new CCSNe models will impact the 
GCE calculations, which require the inclusion of a stellar initial
mass function, star formation history, and other factors.
Therefore, so far, 
the correlations observed in samples of many metal-poor stars 
cannot be explained by current CCSN models.

The abundances reported in the large-sample studies comprise a
heterogeneous
set, with variations in stellar evolutionary states, spectral wavelength
coverage, resolution, and signal-to-noise, and analytical techniques.
Moreover, these abundances may not be reliable due to the lack of accurate
atomic transition data.
In particular, vanadium had fewer reliable  transitions available for measurement prior the
publication of new laboratory atomic data from \citet{lawler14} and \citet{wood14v}.
\citet{roederer14c} presented V~\textsc{i} detections in 134~stars
(including 16 without V~\textsc{ii}) from 1 V~\textsc{i} line, 
V~\textsc{ii} detections in 188~stars (including 70 without V~\textsc{i}) from 2 V~\textsc{ii} lines,
for a total of 204 stars with either V~\textsc{i} or V~\textsc{ii}
detections.
To further test the correlations found by \citet{sneden16},
we apply the improved atomic data to rederive the vanadium abundances for stars in this sample.

In Section~\ref{obs}, we provide a quick review of the high-resolution spectroscopic 
data from \citet{roederer14c}.
We discuss the analysis methods and uncertainty estimates in Section~\ref{methods},
where we also include the individual fitting for each transition and the calibrations applied.
The new vanadium abundances and their correlations with other parameters are presented in
Section~\ref{results}. 
We discuss the significance of our results
in Section~\ref{discussion}, and we conclude in 
Section~\ref{conclusion}.

We adopt the standard definitions of elemental abundances and 
ratios. The logarithmic absolute abundance for element X
is defined as the number of atoms of X per $10^{12}$ hydrogen
atoms, $\log \epsilon\text{(X)} \equiv \log_{10}{(N_{X}/N_{H})} +12.0$.
For elements X and Y, the logarithmic abundance ratio relative to the
solar ratio is defined as 
[X/Y] $\equiv \log_{10}{(N_{X}/N_{Y})} - \log_{10}{(N_{X}/N_{Y})}_\odot$.
We adopt the solar abundances of \citet{asplund09}, where 
$\log \epsilon\text{(V)}=3.93$.
We note that, when denoted with the ionization state (e.g.,
$\log \epsilon\text{(X~\textsc{i})}$),
the value still represents the total elemental abundance, 
derived from transitions of that particular ionization state
after applying Saha ionization corrections.

\section{Observations}
\label{obs}

In this work, we adopt the same sample and spectra used in  
the \citet{roederer14c} study without change.
It should be noted that these targets were selected from various surveys in a highly heterogeneous manner.
They are stars that are deemed to be chemically interesting
and bright enough to be suitable
for high resolution, high S/N spectroscopy. 
Thus, they do not represent an unbiased stellar population.
Yet, the selection criteria do not include any constraints
on the vanadium abundances of the targets, so
this sample was chosen without bias to the vanadium abundances.

For most of the stars, the spectroscopy was obtained 
using the Magellan Telescopes at Las Campanas Observatory
using the Magellan Inamori Kyocera Echelle (MIKE) 
spectrograph \citep{bernstein03}. 
The MIKE spectra cover wavelengths of 3350-7250~\AA, 
with $R \sim 41,000$ and S/N $\sim 40$ per resolution element (RE)
at 3950~\AA.~
Some observations were made with 
the Robert G.\ Tull Coud\'{e} Spectrograph \citep{tull95}
on the 2.7~m Harlan J.\ Smith Telescope at McDonald Observatory (MCD),
with narrower wavelength coverage 3700-5700~\AA, $R \sim 30,000$, 
and S/N $\sim 50$ $\text{RE}^{-1}$ at 3950~\AA.~
The narrower wavelength coverage at the blue end prevents us from using
two useful short-wavelength V~\textsc{ii} lines to derive abundances.
The narrower wavelength coverage at the red end does not cause problems.
Other observations were made with the High Resolution Spectrograph 
on the 9.2m Hobby-Eberly Telescope at McDonald Observatory (HET), 
with wavelength coverage 3900-6800 \AA, $R \sim 30,000$, 
and S/N $\sim 20$ $\text{RE}^{-1}$ at 3950 \AA.~
The 3900~\AA\ blue cutoff of the HET spectra limits the V~\textsc{ii} transitions to
four lines.
Further details of the
instrument setup and observing strategy may be found
in \citet{roederer14c}.

In total, 250 stars were observed with MIKE,
52 with MCD, and 19 with HET.~
Accounting for the repeat observations of stars and 
three double-lined spectroscopic binaries, which are not examined here,
313 stars are available for examination. 
The duplicate observations of the same stars
using different instruments serves as a tool 
to examine consistency across instruments, which is
further discussed in Section~\ref{internal}.

\section{Methods}
\label{methods}

\subsection{Model Atmospheres}
\label{model}

For the model atmospheres,
we adopt the same set of MARCS models \citep{gustafsson08} that 
was used in \citet{roederer14c}.
We assume 1D, plane-parallel, static model atmospheres in local thermodynamic equilibrium (LTE)
throughout the line-forming layers.
As described in \citeauthor{roederer14c},
most of the model atmosphere parameters are derived from the spectra.
The effective temperatures (\teff) are derived
by enforcing that the iron abundances derived from the neutral lines
have no trend with the excitation potential (E.P.) of the lower level
of the transition.
The typical statistical uncertainty in \teff\
ranges from 40-50~K.~
Similarly, microturbulence velocities (\vt) are derived
by requiring no trend with line strength.
The statistical uncertainty in \vt\ is
approximately 0.06~\kmsec.
Surface gravities (\logg) are derived by interpolation from \teff
using theoretical isochrones in the $\text{Y}^{2}$ grid \citep{demarque04},
except for the horizontal branch (HB),
main sequence (MS), and blue straggler-like (BS) classes,
which are derived using the usual method of requiring
the ionization balance between Fe~\textsc{i} and Fe~\textsc{ii}.
The statistical uncertainty in \logg\ ranges between 0.15 and 0.25~dex.
The overall metallicity is simply 
the iron abundance derived from Fe~\textsc{ii} lines.
The statistical uncertainty for metallicity is approximately 0.07~dex.
All uncertainties mentioned above
do not include the systematic uncertainties.
We adopt these derived model parameters 
and the associated models for the spectral synthesis
in this work to derive vanadium abundances.

As discussed extensively in \citet{roederer14c},
systematic uncertainties are very difficult to quantify
accurately.
They can be estimated by comparing the derived 
stellar parameters obtained through different techniques.
\citet{roederer14c} provides  
systematic uncertainties of the stellar parameters
for different evolutionary stages 
in this sample.
For example, the systematic difference in \logg\ values
can be calculated from the $\text{Y}^{2}$ set of isochrones 
and another set in common use, 
the PARSEC grid \citep{bressan12}.
For metal-poor red giants, the difference in \logg\  
at fixed \teff\ is $\approx$~0.2--0.3~dex,
with the $\text{Y}^{2}$ set of models giving larger \logg\ values.
For metal-poor subgiants, the difference is even smaller, 
$<$~0.1~dex, and the two sets of isochrones 
yield nearly identical results around 6000~K.
These systematic differences are comparable to the 1$\sigma$ 
statistical uncertainties in the \logg\ values, 
and they would affect all stars of similar \teff\ equally.
Such changes have minimal impact on the derived [V/Fe] ratios.
(Section~\ref{trendsp}).
For our purpose, which is to compare abundances ratios
of stars within the sample itself,
it is sufficient to just consider the 
statistical uncertainties.
Furthermore, when we discuss the correlation of vanadium abundances
with other iron-group elements in Section~\ref{correlation},
the abundance ratios of those elements are also derived 
based on the same models.

\subsection{Abundance Analysis}
\label{analysis}

We derive the vanadium abundances using synthetic spectra generated with 
the LTE line analysis code MOOG \citep{sneden73,sobeck11}.
The analysis code produces synthetic spectra sets based on different
assumptions on vanadium abundance of each star.
We then obtain the vanadium abundance for each star
by iterative comparison of the synthetic spectra to the observed spectrum.
Line lists used for the syntheses, including the new laboratory transition 
data for both V~\textsc{i} \citep{lawler14} and V~\textsc{ii} \citep{wood14v},
are listed in Table~\ref{linetab}.
We start with 6 V~\textsc{i} and 20 V~\textsc{ii} lines
that are available in the wavelength range of our spectra.
We generate test syntheses for all 26~lines in
stars ranging from metal-poor subgiants (CS~22886--012)
to metal-rich red giants (HD~21581).
We identify 4 V~\textsc{i} and 7 V~\textsc{ii} lines 
that are relatively strong
and unblended across a range of \teff\ and metallicities.
The table provides the E.P.\ and 
the \loggf\ value for each line we select.
There are more lines available as compared to those in \citet{roederer14c}.
All of them, except for one V~\textsc{i} line and one V~\textsc{ii} line, 
include hyperfine structure (HFS).
They are indicated in Table~\ref{linetab}.

The HFS line component patterns for V~\textsc{i} have been constructed from
laboratory measurements of the HFS $A$ (magnetic dipole) constants, 
and (much smaller) $B$ (electric quadrupole) constants when available, 
as discussed in \citet{lawler14} and references therein.
\citeauthor{lawler14}\ combined previous measurements of HFS $A$ constants 
with their FTS spectra to extract HFS $A$ constants for many additional levels
that had not been measured directly, 
and we make use of the expanded set of transitions connecting these levels.
\citet{wood14v} used a similar strategy to measure new HFS $A$ constants 
for V~\textsc{ii}.
These patterns are of high quality and permit reliable de-saturation of 
strong V~\textsc{i} and V~\textsc{ii} lines in stellar spectra.
Lines with no available HFS are marked in Table~\ref{linetab}.
These lines had no or minimal profile broadening in the laboratory 
spectra, indicating that their HFS constants are small,
so they can be treated as single lines with minimal loss of accuracy.

\startlongtable
\begin{deluxetable}{cccc}
\tablecaption{Transitions of Neutral and Ionized Vanadium
\label{linetab}}
\tablewidth{0pt}
\tabletypesize{\scriptsize}
\tablehead{
\colhead{Species} &
\colhead{$\lambda$ ( \AA )} &
\colhead{E. P. (eV)} &
\colhead{\loggf}
}
\startdata
  V~\textsc{i}   & 4111.779                  &      0.30     &      $+$0.40 \\
  V~\textsc{i}   & 4379.230\tablenotemark{a} &      0.30     &      $+$0.58 \\
  V~\textsc{i}   & 4389.979\tablenotemark{b} &      0.28     &      $+$0.22 \\
  V~\textsc{i}   & 4408.196                  &      0.28     &      $-$0.05 \\
\hline
 V~\textsc{ii}   & 3517.299                  &      1.13     &      $-$0.24 \\
 V~\textsc{ii}   & 3545.196                  &      1.09     &      $-$0.32 \\
 V~\textsc{ii}   & 3715.464\tablenotemark{b} &      1.57     &      $-$0.22 \\
 V~\textsc{ii}   & 3951.957\tablenotemark{a} &      1.48     &      $-$0.73 \\
 V~\textsc{ii}   & 4002.928                  &      1.43     &      $-$1.44 \\
 V~\textsc{ii}   & 4005.702\tablenotemark{a} &      1.82     &      $-$0.45 \\
 V~\textsc{ii}   & 4023.377                  &      1.80     &      $-$0.61 \\
\enddata      
\tablenotetext{a}{Lines also used in \citet{roederer14c}.}
\tablenotetext{b}{Lines without HFS.}
\end{deluxetable}

When fitting the synthetic spectra to the observed spectrum, 
in most cases we adjust only the smoothing.
For some of the evolved red giants with \teff\ lower than 5200~K,
where molecular CH bands are significant, 
we adopt a typical value of $^{12}$C/$^{13}$C = 5 (e.g., \citealt{gratton00})
when generating the synthetic spectra. 
Doing so greatly improves the agreement between synthetic and observed 
spectra near the CH band region, and thus
provides a better fit for the V~\textsc{i} line 
at 4389~\AA.~
There are 102 stars with synthetic spectra configured this way,
most of which are red giants. 
The adopted $^{12}$C/$^{13}$C has no impact on the derived V abundances for all other stars
or any other V~\textsc{i} or V~\textsc{ii} lines.

\begin{figure*}
\begin{center}
\includegraphics[angle=0,width=6.85in]{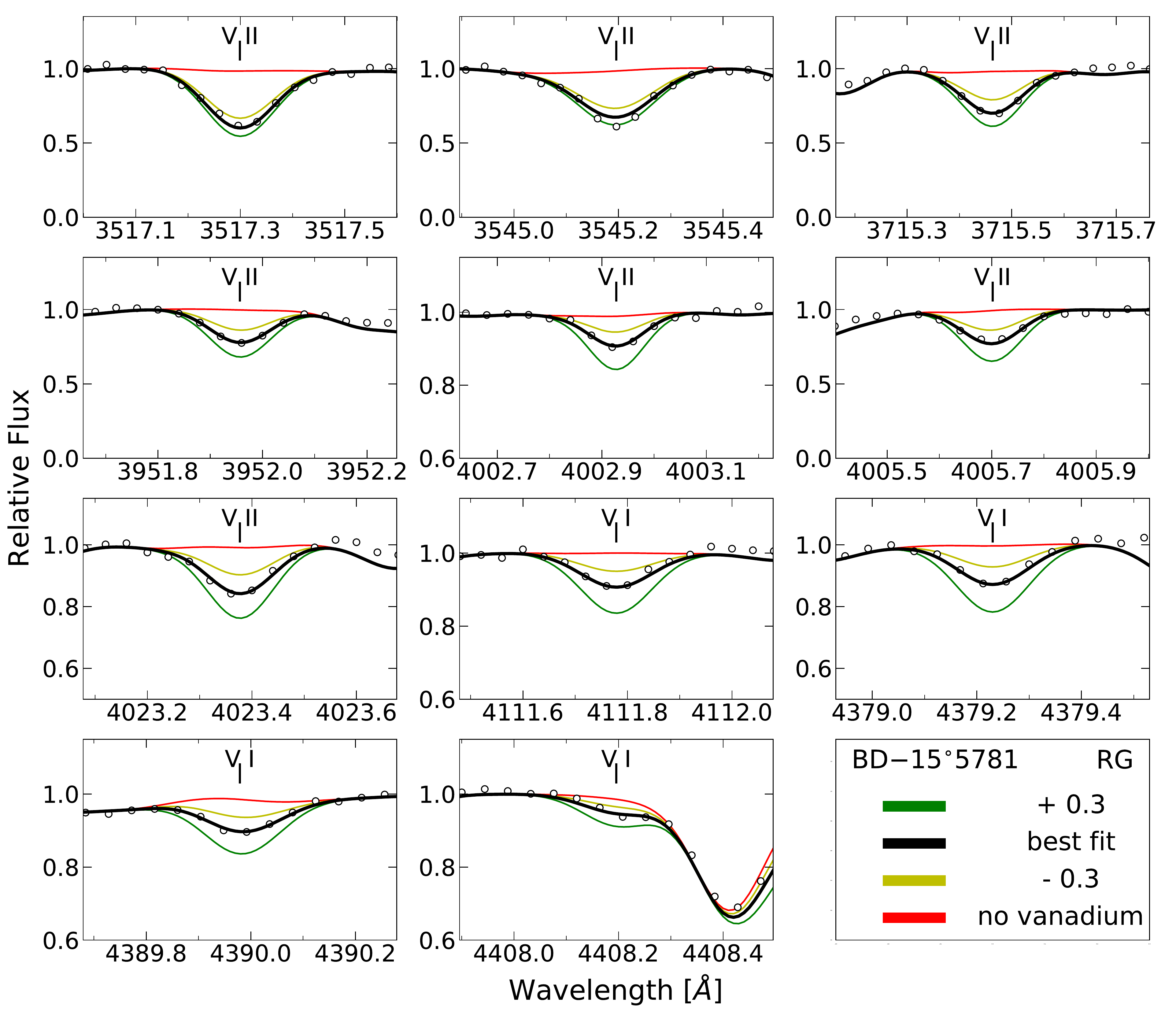}
\end{center}
\caption{
Observed (points) and synthetic (lines) spectra
for a representative red giant, \mbox{BD~$-$15\degree5781}.
The first seven lines shown are V~\textsc{ii} lines,
while the last four lines are V~\textsc{i} lines.
The best-fit lines are for each line individually.
Note that the vertical axis range is adjusted 
for some of the weak lines to provide a clearer visualization
of the best fit.
}
\label{rgmoogfit}
\end{figure*}

\begin{figure*}
\begin{center}
\includegraphics[angle=0,width=6.85in]{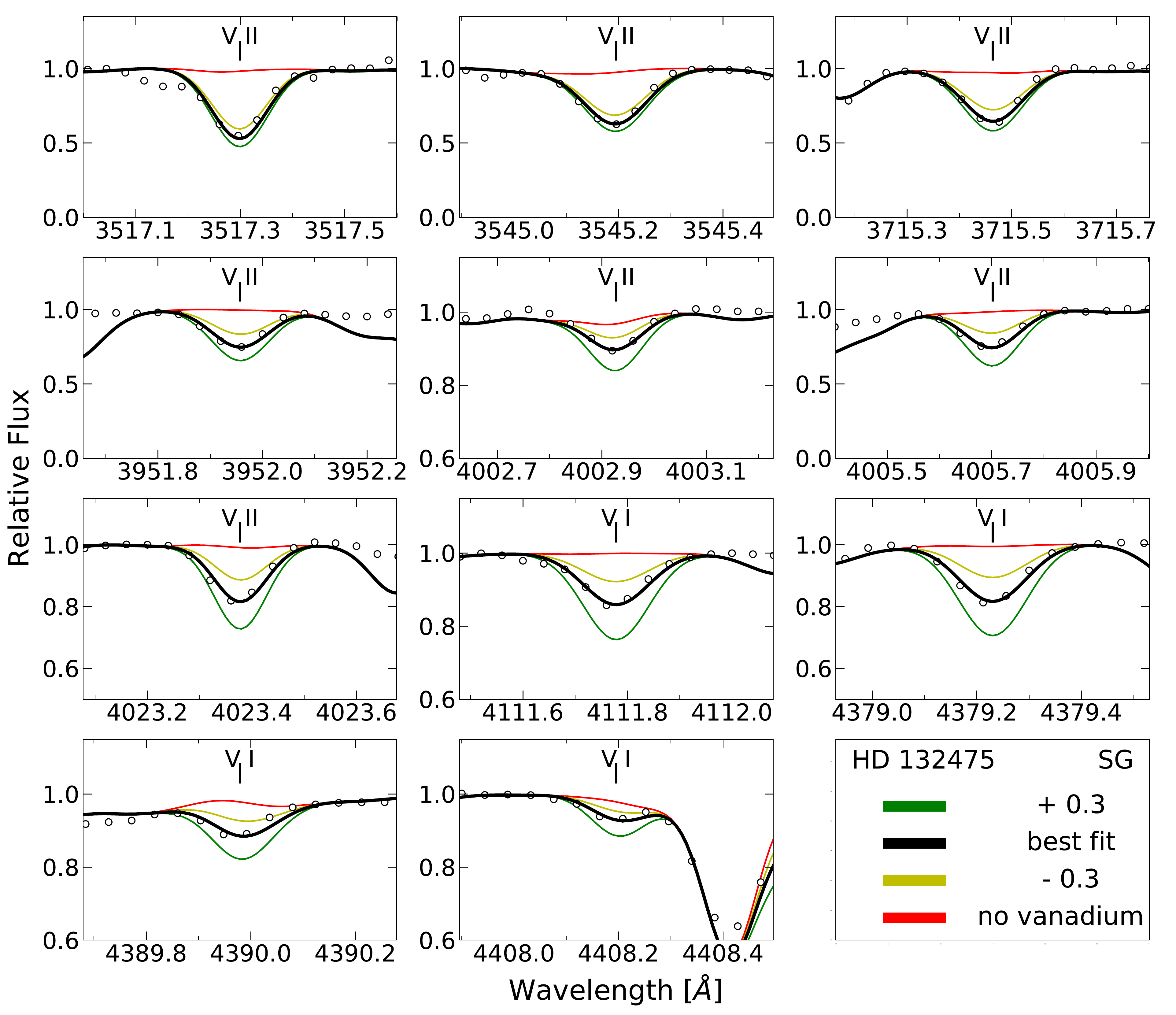}
\end{center}
\caption{
Observed (points) and synthetic (lines) spectra for a representative sub-giant, HD~132475.
}
\label{sgmoogfit}
\end{figure*}

Figure~\ref{rgmoogfit} and Figure~\ref{sgmoogfit} present two synthetic matching examples
with \mbox{BD~$-$15\degree5781} (a red giant) 
and HD~132475 (a sub-giant).
The resulting line-by-line $\log \epsilon$ measurements
are summarized in Tables~\ref{v1tab} and~\ref{v2tab},
where we include the wavelength ($\lambda$) at which the
measurement is made. 
In Table~\ref{vtab}, we also present the result combining all measurements
for each star, together with the class, effective temperature (\teff),
surface gravity (\logg), as well as metallicity ([Fe~\textsc{i}/H] and [Fe~\textsc{ii}/H]).
For stars with only an upper limit available,
we indicate that fact with an upper limit flag (U. L.),
and we note the line from which the upper limit is determined.

\startlongtable
\begin{deluxetable}{cccc}
\tablecaption{Abundances Derived from Neutral Vanadium Lines
\label{v1tab}}
\tablewidth{0pt}
\tabletypesize{\scriptsize}
\tablehead{
\colhead{Star} &
\colhead{Wavelength (\AA)} &
\colhead{$\log \epsilon$ (V~\textsc{i})} &
\colhead{$\sigma$}
}
\startdata
      BD~$+$10\degree2495 &        4111.779 &            1.51 &            0.10 \\ 
      BD~$+$10\degree2495 &        4379.230 &            1.39 &            0.06 \\ 
      BD~$+$10\degree2495 &        4389.979 &            1.42 &            0.10 \\ 
      BD~$+$10\degree2495 &        4408.196 &            1.33 &            0.20 \\ 
     BD~$+$19\degree1185a &        4111.779 &            2.44 &            0.20 \\ 
     BD~$+$19\degree1185a &        4379.230 &            2.53 &            0.10 \\ 
     BD~$+$19\degree1185a &        4389.979 &            2.56 &            0.15 \\ 
     BD~$+$19\degree1185a &        4408.196 &            2.38 &            0.10 \\ 
\enddata      
\tablecomments{%
The model atmosphere uncertainty is not included in here for reasons 
discussed in Section~\ref{uncertainties}.
The complete version of Table~\ref{v1tab} is available in the online edition of the journal. 
A short version is included here to demonstrate its form and content.
}
\end{deluxetable}

\startlongtable
\begin{deluxetable}{cccc}
\tablecaption{Abundances Derived from Ionized Vanadium Lines
\label{v2tab}}
\tablewidth{0pt}
\tabletypesize{\scriptsize}
\tablehead{
\colhead{Star} &
\colhead{Wavelength (\AA)} &
\colhead{$\log \epsilon$ (V~\textsc{ii})} &
\colhead{$\sigma$}
}
\startdata
      BD~$+$10\degree2495 &        3715.464 &            1.70 &            0.17 \\ 
      BD~$+$10\degree2495 &        3951.957 &            1.77 &            0.18 \\ 
      BD~$+$10\degree2495 &        4002.928 &            1.69 &            0.11 \\ 
      BD~$+$10\degree2495 &        4005.702 &            1.87 &            0.15 \\ 
      BD~$+$10\degree2495 &        4023.377 &            1.87 &            0.11 \\ 
     BD~$+$19\degree1185a &        3715.464 &            2.74 &            0.18 \\ 
     BD~$+$19\degree1185a &        3951.957 &            2.76 &            0.14 \\ 
     BD~$+$19\degree1185a &        4002.928 &            2.71 &            0.20 \\ 
     BD~$+$19\degree1185a &        4005.702 &            2.86 &            0.06 \\ 
\enddata      
\tablecomments{%
The model atmosphere uncertainty is not included in here for reasons 
discussed in Section~\ref{uncertainties}.
The complete version of Table~\ref{v2tab} is available in the online edition of the journal. A short version is included here to demonstrate its form and content.
}
\end{deluxetable}

\subsection{Uncertainties}
\label{uncertainties}

The statistical uncertainty for $\log \epsilon$(V) of each line includes
the fitting uncertainty of the measurement and the uncertainty in the \loggf\ 
of the line transition \citep{lawler14,wood14v}. 
The fitting uncertainties are estimated based on the 
spectrum synthesis matching.
The typical fitting uncertainty for the sample 
is approximately 0.15~dex.
In most cases, the fitting uncertainty is much larger 
than the transition probability uncertainty,
because the lines we adopted are all major branches
that are understood and measured to great precision (typically 0.02~dex uncertainty)
in the laboratory.
For each given star,
we weight the vanadium abundance measurement from
each line by the inverse-square of the statistical uncertainty
to calculate the total statistical uncertainty of the star.
In cases when the few measurements agree exactly,
we set an arbitrary minimum statistical weighted uncertainty of 0.05~dex.

The abundance uncertainty introduced by the uncertainties 
in the model atmosphere parameters
is extracted from the vanadium abundances
derived by \citet{roederer14c}.
Although we use more lines in this work,
we still decided to adopt the same model atmosphere uncertainties.
While individual lines with different E.P. will have different model
\teff\ sensitivities,
the lines we have used here have a very similar range of E.P.\ as the ones 
used in \citet{roederer14c}, as shown in Table~\ref{linetab}.
Thus we expect the newly added lines will respond to the model
atmosphere parameter uncertainties similarly to the lines
used in the previous work.
We combine the extracted model atmosphere uncertainties with
the total fitting statistical uncertainties. 

Our study has made several improvements relative to \citet{roederer14c}
that reduce the overall abundance uncertainties in most stars.
In particular, we use improved atomic transition data with almost 
negligible uncertainties in the \loggf\ value of the line.
More importantly, we use more lines so that we have better statistics. 
The statistical uncertainty approaches the fitting uncertainty
as the number of lines we use for each star increases,
assuming there is no intrinsic scattering in the vanadium 
measurement from different lines.
The typical size of this statistical uncertainty is 0.10 for the
50 most metal-poor stars and 0.08 for the 50 most metal-rich stars.
Yet, the final total uncertainties are dominated by the
model uncertainty because the typical size of the model atmosphere 
uncertainties is approximately 0.17.
On average, the total uncertainties increase by only a small amount 
($< 10$\%) as we move from
the metal-rich to the metal-poor stars.

\subsection{Internal Comparisons}
\label{internal}

Stars with just one measured 
V~\textsc{i} or V~\textsc{ii} transition
have larger star-to-star scatter in a given [Fe/H] domain than do stars
with multiple measurements.  
We perform statistical tests to determine
if single-line abundance results should be dropped from further analyses.
Statistical tests (KS test, \citealt{kolmogorov33,smirnov48}; 
Anderson test, \citealt{stephens74}; 
Mann-Whitney U test, \citealt{mann47}) 
show a strong indicator that the stars with only one measurement 
are not drawn from the same distribution as the stars with multiple measurements.
The KS test, for example, gives a p-value of $1.15\times 10^{-9} $.
We thus treat these stars with less confidence
and advise readers to be cautious
when using these measurements for other purposes.
In most of the figures, we use ``x'' s 
to indicate the stars with only one measurement,
and we exclude these stars from further analysis of the aggregated sample.

\begin{figure}
\begin{center}
\includegraphics[angle=0,width=3.35in]{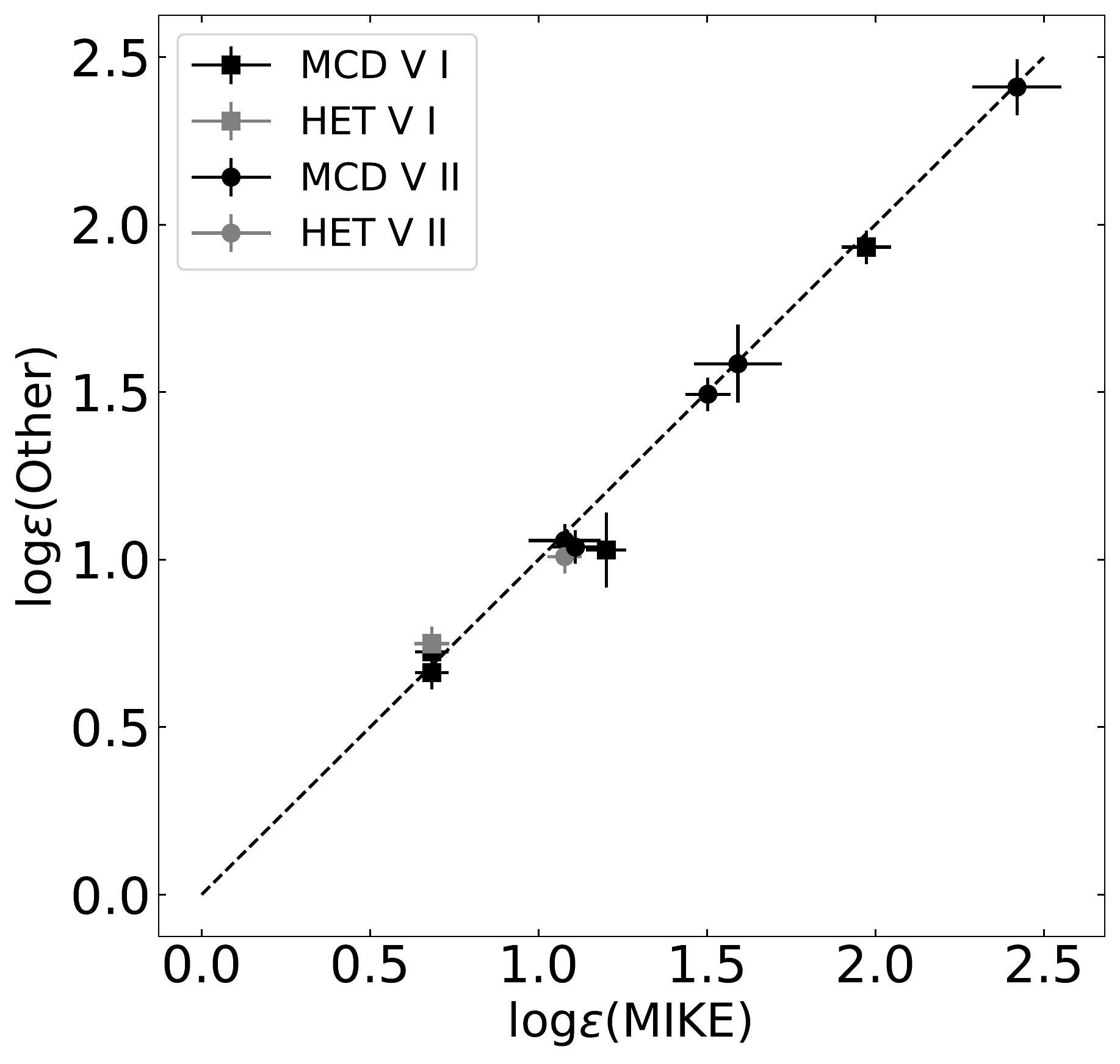}
\end{center}
\caption{
\label{common}
Comparison of the stars observed with multiple spectrographs
in the sample. 
Each data point represents a star, and
the position of the data point represents 
their derived $\log \epsilon$ value from different spectra.
The black dots give the comparison between MIKE and MCD,
and the gray square gives the comparison between MIKE and HET.~
The dashed line represents perfect agreement.
 }
\end{figure}

To examine the consistency between 
spectra taken by different telescopes and instruments,
we examined the five stars that are observed multiple times 
with MIKE, MCD, and HET:\
HD~106373,
HD~108317,
HD~122563,
HD~132475,
and HE~1320-1339.
All of these stars have one observation in MIKE,
and another observation in MCD.~
HD~122563 was also observed with the HET.~
We find good agreement, both through a line-by-line abundance comparison
and a weighted average abundance comparison.
As shown in 
Figure~\ref{common}, most of the results agree
to within 1$\sigma$,
which we consider to be satisfactory.

\subsection{Comparison with Previous Results}
\label{previous}

\begin{figure*}
\begin{center}
\includegraphics[angle=0,width=2.5in]{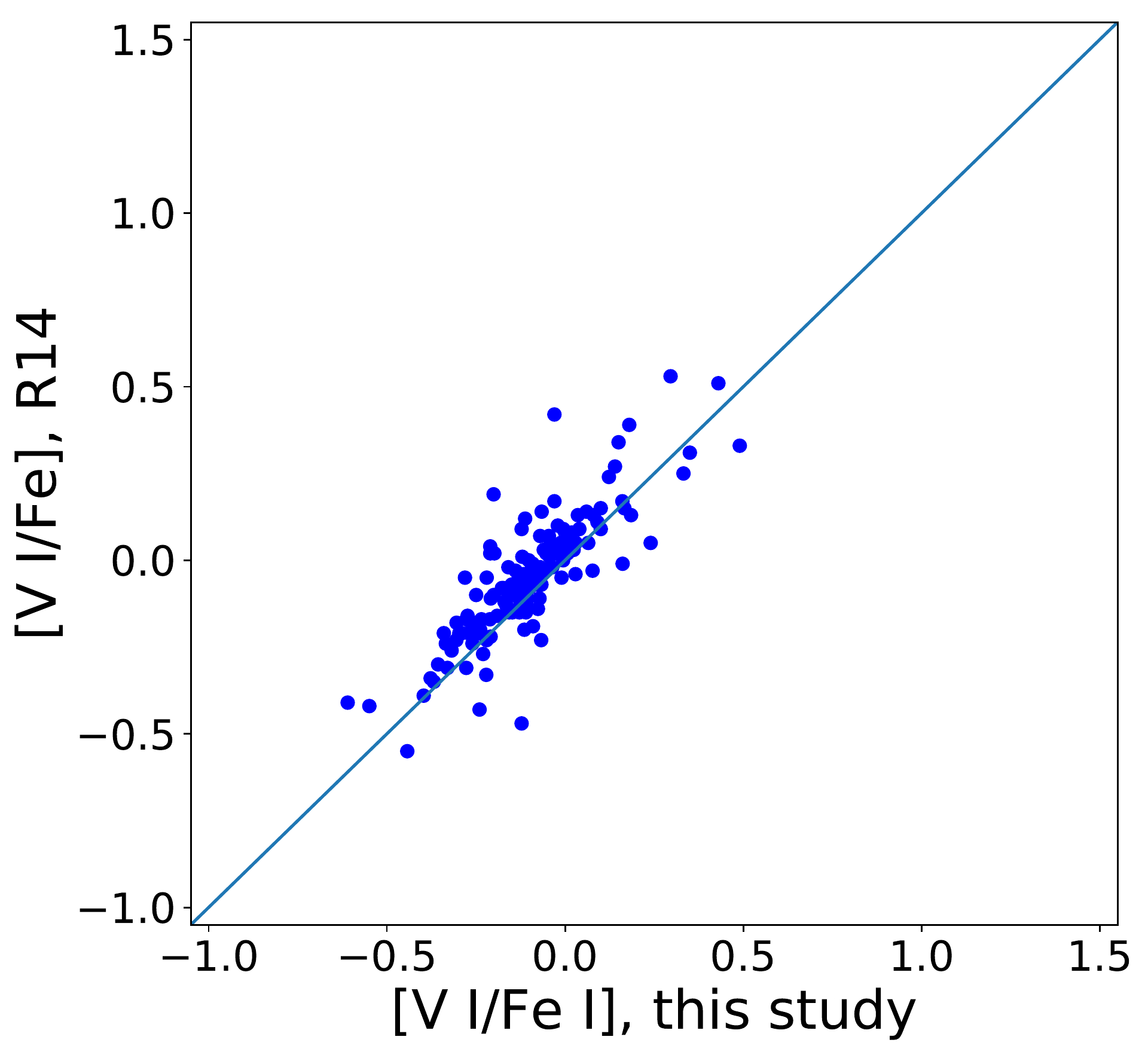}
\hspace{0.02in}
\includegraphics[angle=0,width=2.5in]{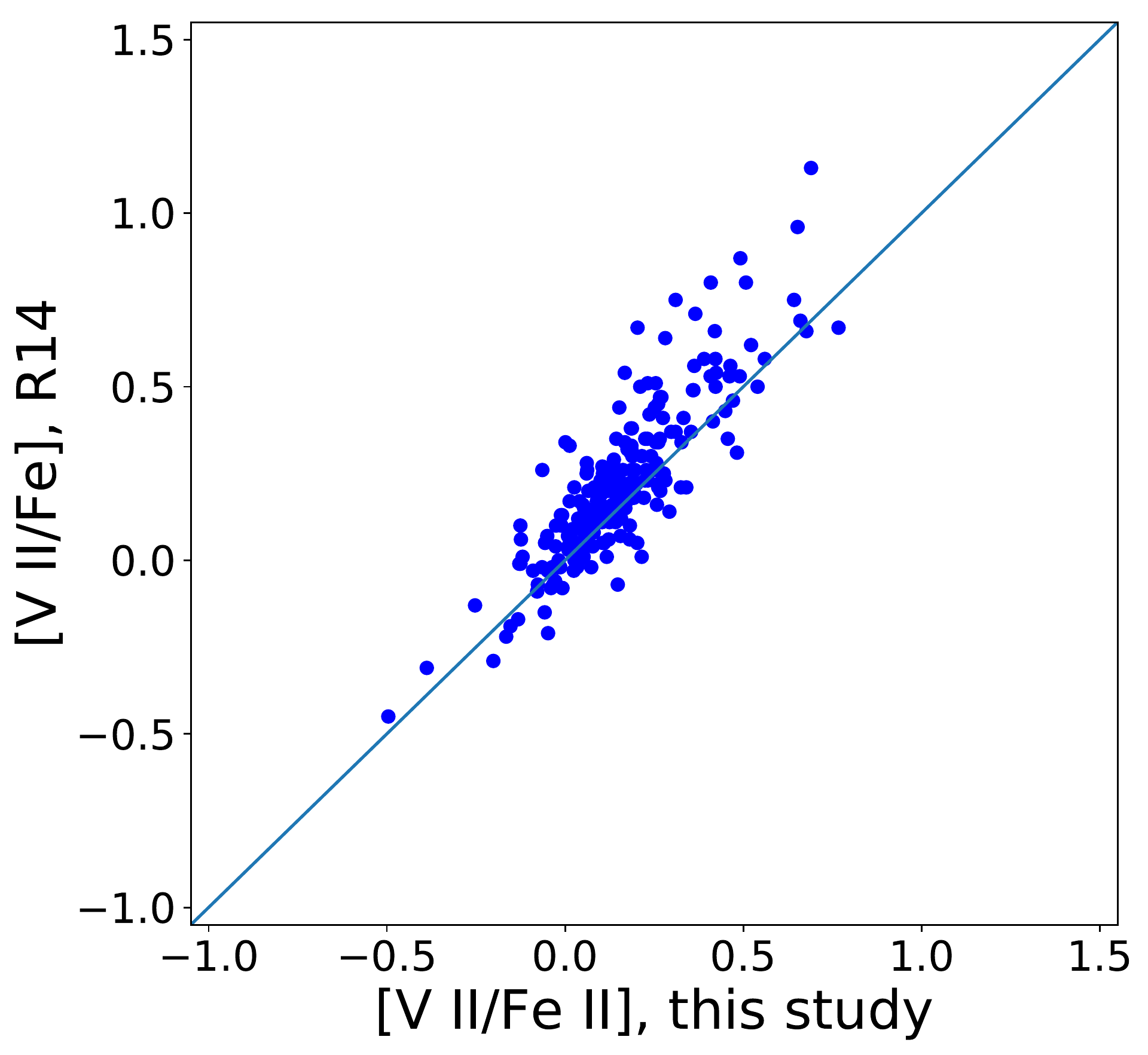} \\
\includegraphics[angle=0,width=2.5in]{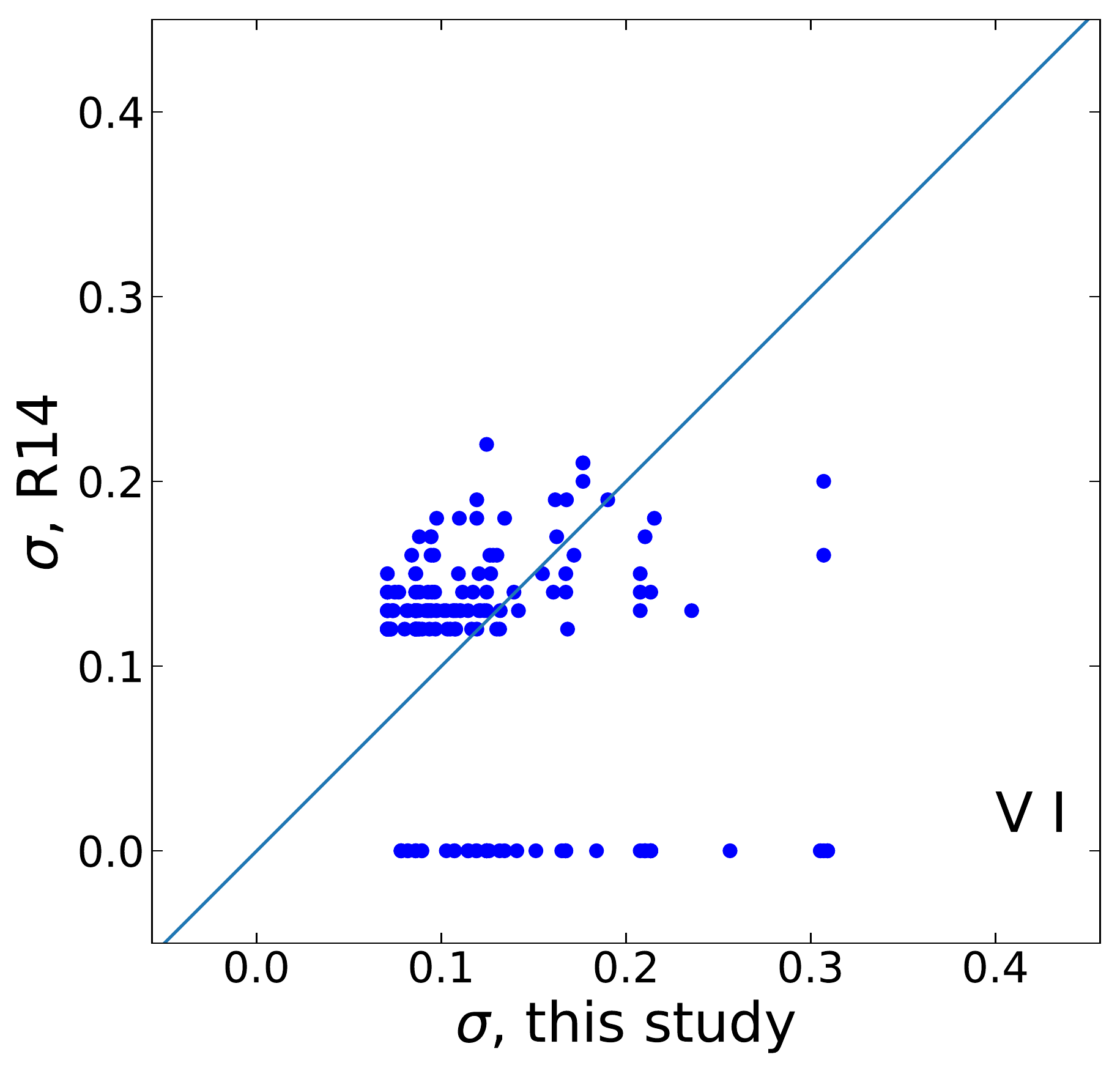}
\hspace{0.02in}
\includegraphics[angle=0,width=2.5in]{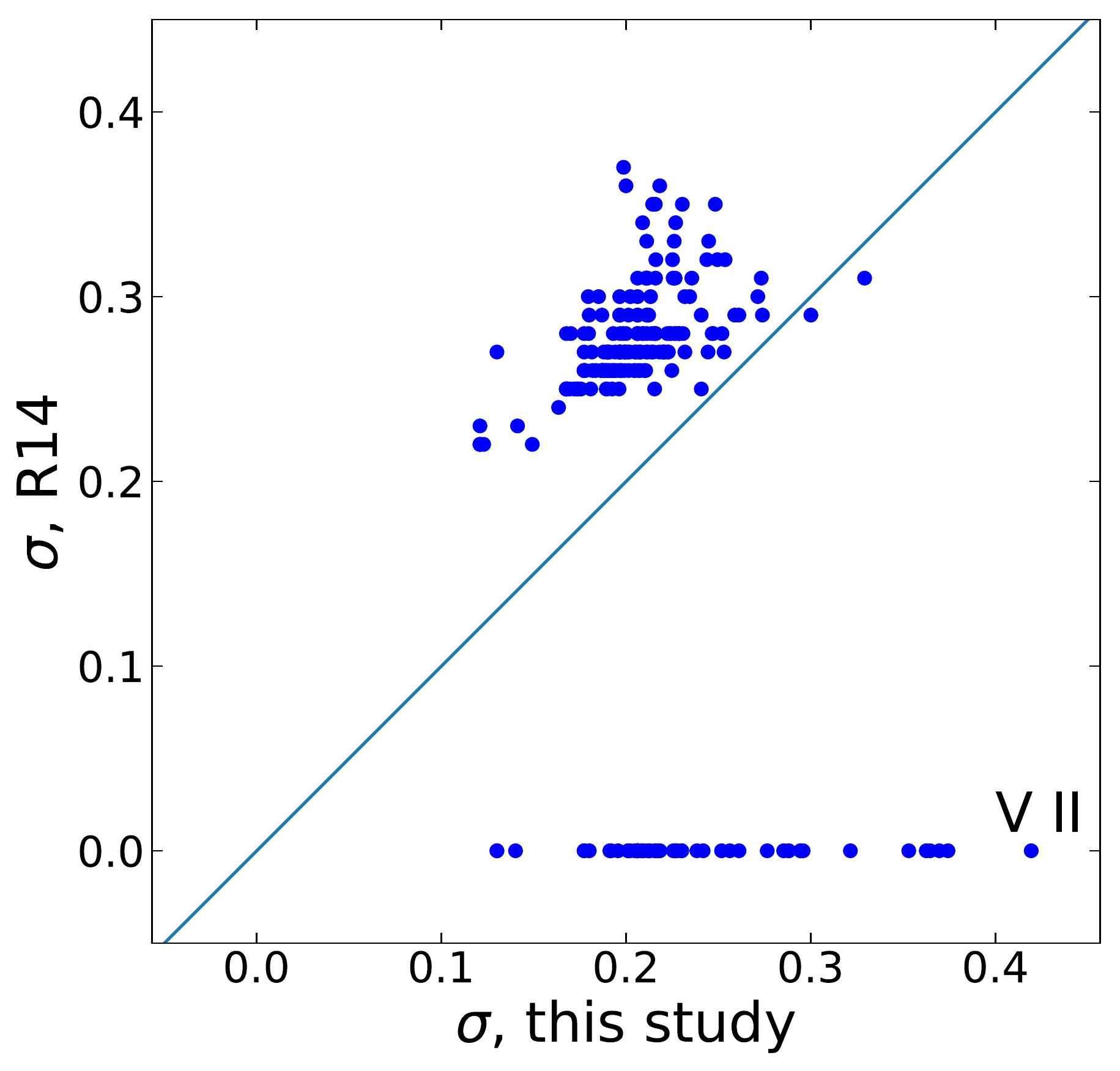}
\end{center}
\caption{
\label{com_iur14}
Comparison of the [V/Fe] ratios and
the uncertainties with those
from \citet{roederer14c} (R14).
The blue lines represents perfect agreement between
the two studies.
The points at y=0.0 in the $\sigma$ plots correspond to 
stars with no measurement of V in the R14 sample.
 }
\end{figure*}

\begin{figure}
\begin{center}
\includegraphics[angle=0,width=3.35in]{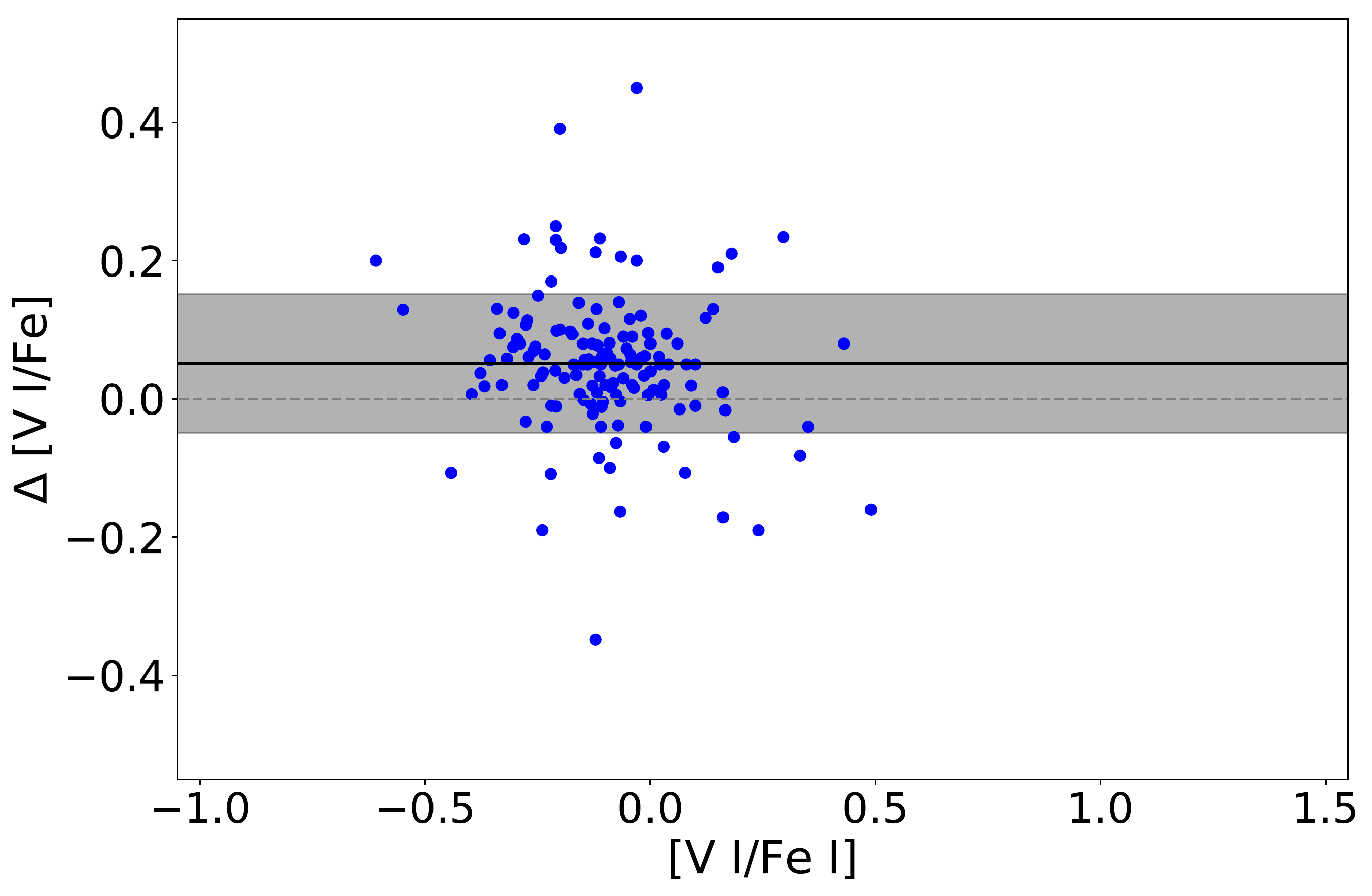} \\
\includegraphics[angle=0,width=3.35in]{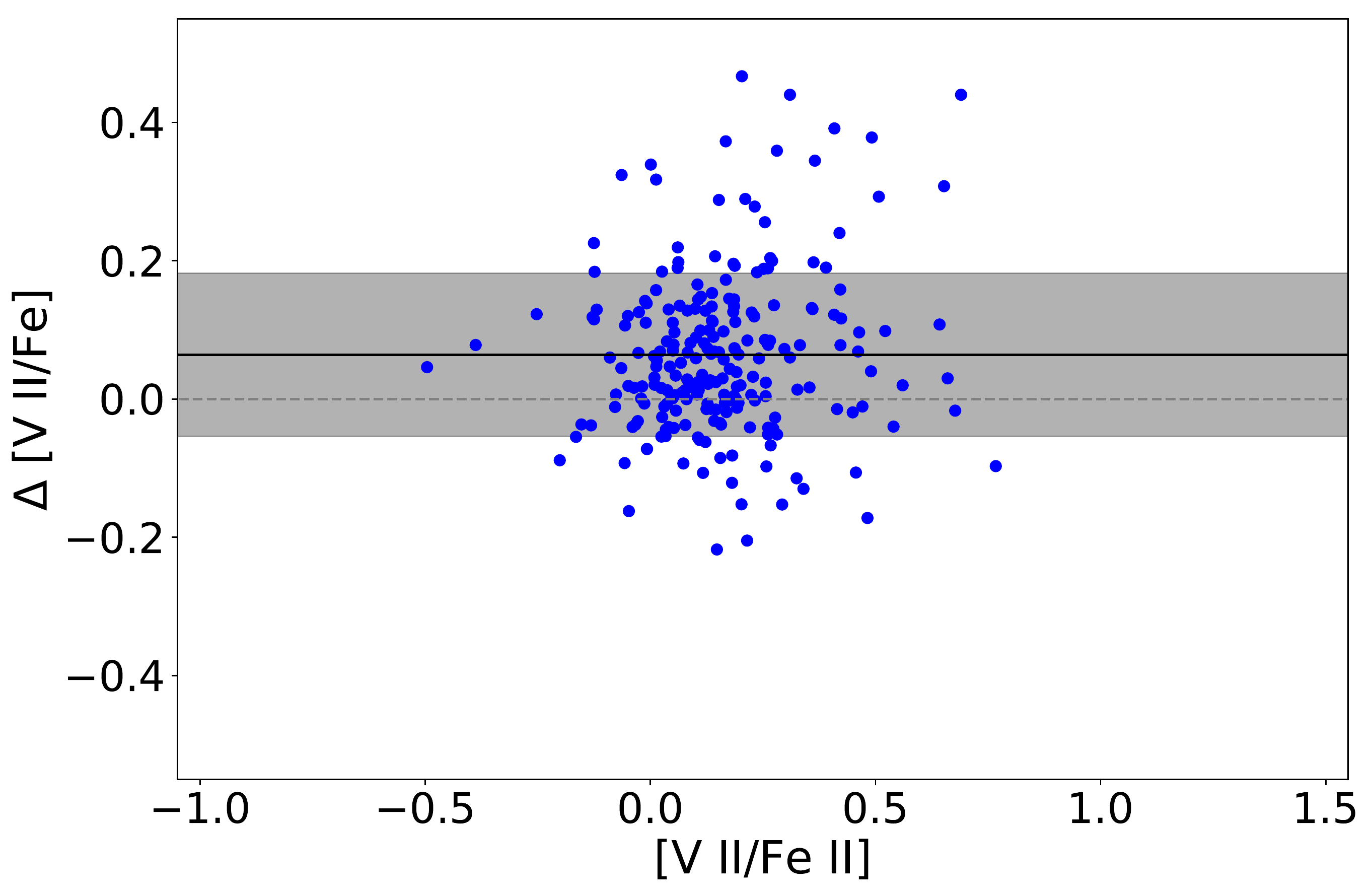}
\end{center}
\caption{
\label{com_iur14_delta}
The difference between this study and \citet{roederer14c}
as a function of the [V/Fe] ratios derived in this work.
The dashed line represents no difference. 
The solid black line and the gray band represent the 
mean and 1$\sigma$ interval of the differences.
 }
\end{figure}

We compare 
our derived $\log \epsilon$(V) values to those
obtained previously
by \citet{roederer14c}.
We find in general good agreement between them
as shown in the top panels of Figure~\ref{com_iur14}.
The offsets remain consistent 
with zero within 1$\sigma$ as shown in Figure~\ref{com_iur14_delta}.
Furthermore, the remaining offsets (0.05~dex for V~\textsc{i} 
and 0.06~dex for V~\textsc{ii})
can be entirely explained by the 
differences in the \loggf\ values adopted by each study.
We also examine the uncertainties 
in these values.
The bottom panels of Figure~\ref{com_iur14} compare the uncertainties
derived by \citet{roederer14c} with those derived in our work.
As expected, the uncertainties are in general improved,
especially for 
abundances derived from V~\textsc{ii} lines,
thanks to the improved atomic transition data and increased
number of lines used.
Comparing the stars that have measurements in both
studies, we improve the median uncertainties
in the logarithmic abundances
by 27~\% (from 0.13 to 0.09~dex) for V~\textsc{i} and 
26~\% (from 0.27 to 0.20~dex) for V~\textsc{ii}.

We also compared our result with the 
metal-poor star
HD~84937 studied extensively in 
\citet{lawler14}, \citet{wood14v}, and
\citet{sneden16}.
Using spectra that cover a wider wavelength range,
those studies provided detailed chemical abundances
for this star with updated transition data,
which we have also used in this work.
As a consistency check,
we adopt the
ATLAS9 AODFNEW model with stellar parameters 
adopted by these previous studies
(\teff\ = 6300~K, \logg\ = 4.0, [Fe/H] = $-2.15$, \vt\ = 1.5~\kmsec).
We obtain individual line measurements that 
agree with those from \citet{lawler14} for V~\textsc{i} 
and \citet{wood14v} for V~\textsc{ii} with differences $\leq$~0.03~dex, 
which is well within our measurement uncertainty. 
For mean abundances, \citet{sneden16} reported 
$\log \epsilon (\text{V}~\textsc{i}) = 1.890 \pm 0.070$ from 9~lines
and $\log \epsilon (\text{V}~\textsc{ii}) = 1.871 \pm 0.075$ from 68~lines. 
After applying the same calibration applied to all other stars in our sample (see Section~\ref{trendwl}), 
we obtain $\log \epsilon (\text{V}~\textsc{ii})$ = 1.95, giving [V~\textsc{ii}/Fe~\textsc{ii}] $= +0.33$. 
For V~\textsc{i}, we detect only the line at 4379~\AA.~
The single measurement, $\log \epsilon (\text{V}~\textsc{i}) = 1.87$, gives [V~\textsc{i}/Fe~\textsc{i}] = +0.26.
Both agree with the \citet{sneden16} values.

\subsection{Trends with Wavelength}
\label{trendwl}

\begin{figure*}
\begin{center}
\includegraphics[angle=0,width=4.7in]{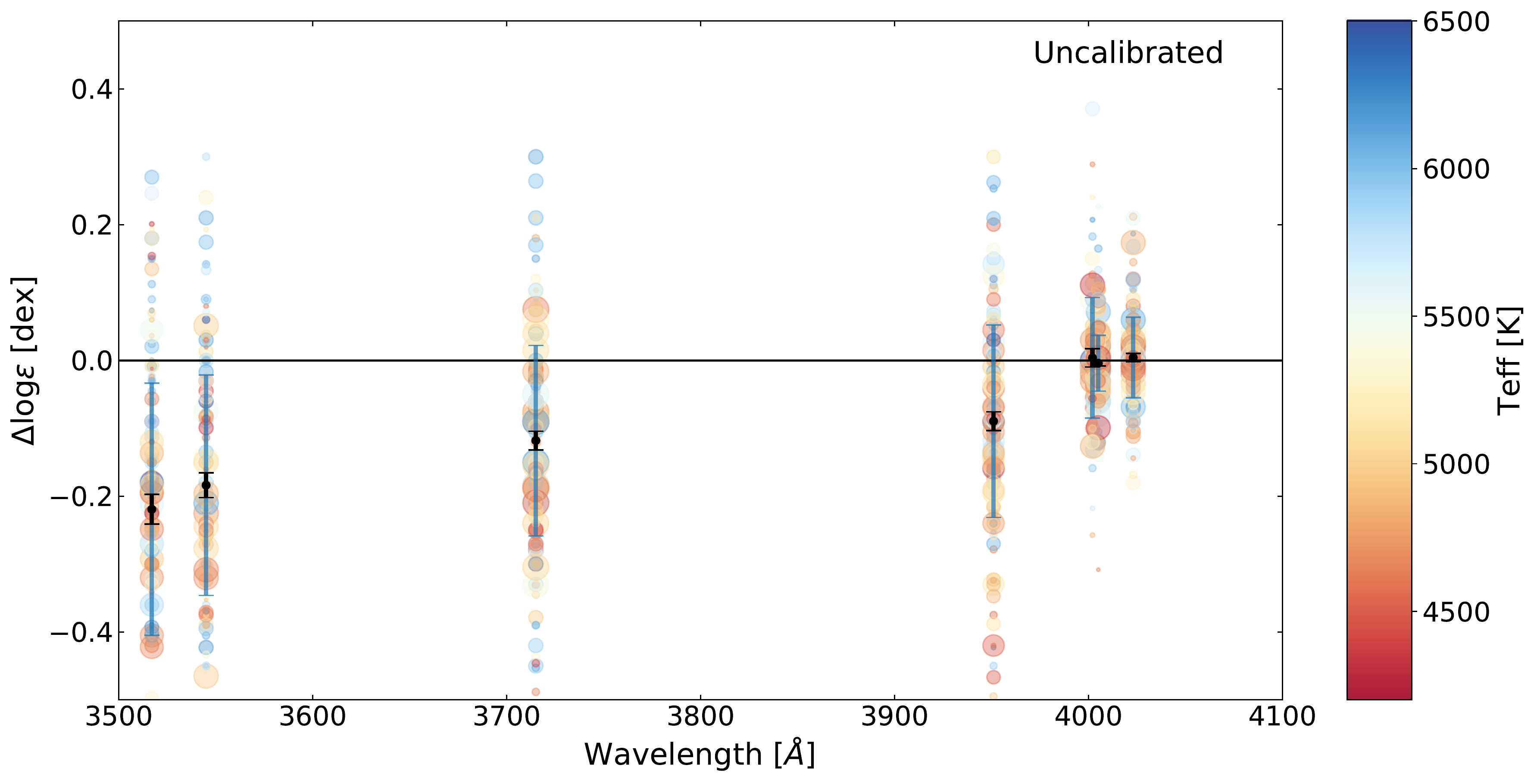} \\
\includegraphics[angle=0,width=4.7in]{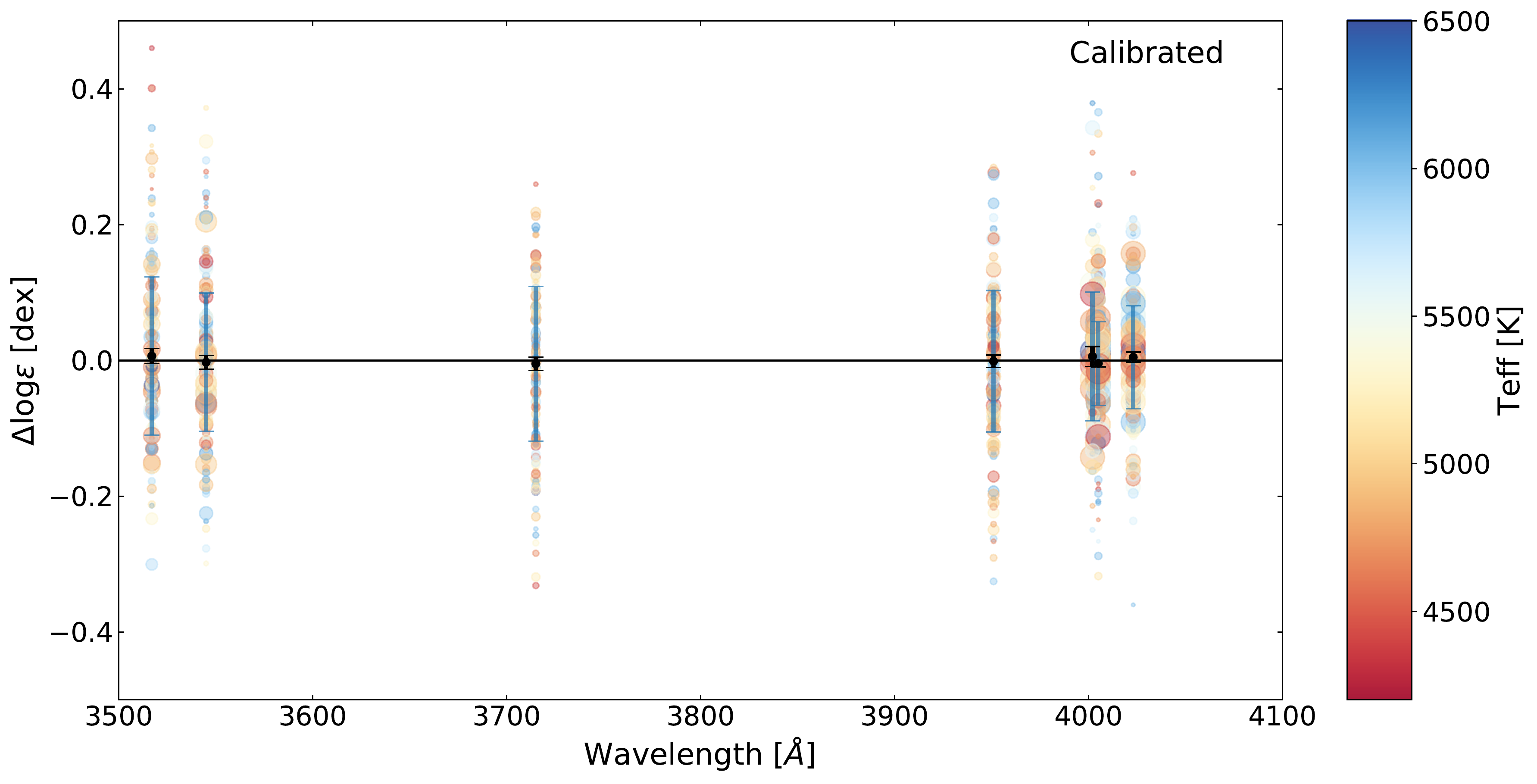}
\end{center}
\caption{
\label{caliplot}
The difference between the fitting
from an individual line and the average calculated from the three
longest wavelength lines,
averaged over all stars
with MIKE data, 
before calibration 
(top panel)
and after calibration
(bottom panel).
The horizontal line represents no offset relative 
to the mean abundance derived from the three
lines with $\lambda > 4000$~\AA.~
The data points are color coded by temperature, 
with point size determined by the uncertainty in the measurement.
 }
\end{figure*}

As discussed extensively in 
\citet{roederer12d,roederer14c,roederer18b}, 
the continuous opacity is mainly contributed by the H$^{-}$\ in the 
atmosphere for most of the stars in our sample, 
with a minor contribution from the bound-free opacity from H~\textsc{i}.
At wavelengths shorter than the Balmer limit at $\approx 3647$~\AA,
the bound-free opacity increases from the Paschen continuum to
the Balmer continuum with a discontinuous jump,
causing an overall increase in the overall continuous opacity.
This change in the continuous opacity may not be fully and accurately accounted for in
the spectrum synthesis, possibly leading to underestimated results for the
vanadium abundances from the four shortest wavelength lines 
(3517.299~\AA, 3545.196~\AA, 3715.464~\AA, and 3951.957~\AA), 
as shown in Figure~\ref{caliplot}.
Neglecting 3D convection effects could also lead to a similar result, 
which explains why some species experience this effect 
but others may not 
(see discussion in \citealt{roederer18b}).
Underestimated \logg\ values 
will not explain the low abundances
from lines at wavelengths shorter than the Balmer limit
because changing changing the \logg\ values of the model
atmosphere would similarly affect the abundances derived from
all V~\textsc{ii} lines examined in this study.

This so-called ``Balmer Dip Effect'' is
significant for about half of the available V~\textsc{ii} lines.
We decided it is 
necessary to apply an empirical calibration to the four shortest wavelength 
V~\textsc{ii} lines. 
The calibration is determined based on the weighted average 
$ \log \epsilon $(V~\textsc{ii}) for each star 
computed with only the three longest wavelength lines
(4002.928~\AA, 4005.702~\AA, and 4023.377~\AA),
which we have taken to be the correct vanadium abundance and are
not affected by the continuum change. 

\begin{figure*}
\begin{center}
\includegraphics[angle=0,width=4.7in]{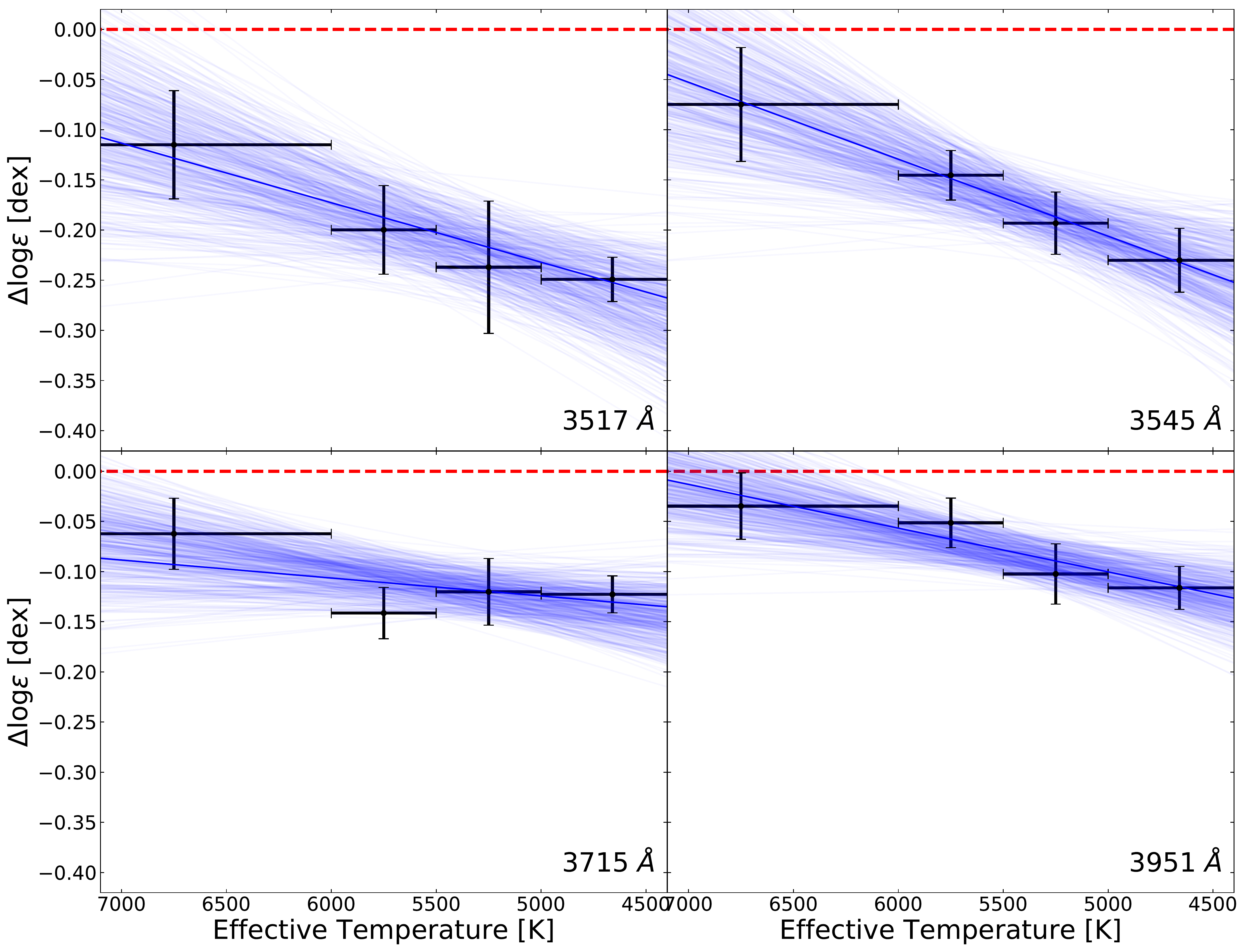} 
\end{center}
\caption{
\label{califit}
The calibration fitting for the four bluest V~\textsc{ii} lines
that are affected by the Balmer region continuum effect.
The horizontal error bars represent the range of \teff\ of
the stars that are used to generate the data point.
The vertical error bars represent the uncertainty in the 
offset that needs to be calibrated for the particular \teff\ group.
The set of lines in the figure is generated using a Monte Carlo
method to provide a visualization of the fitting uncertainty.
They are generated by fitting the resampled offsets from
the four \teff\ groups.
 }
\end{figure*}

To keep consistency in how each star in the sample contributes to determining 
the calibration, we assessed the calibration using only the stars observed with
MIKE, as only they have spectra extended across all seven lines.
This effect is most significant in cooler stars, and it gets weaker as the 
temperature increases. 
We separate the MIKE stars into four temperature groups and fit a line 
to determine the amount of calibration needed for each line
as a function of temperature.
The fitting and its corresponding uncertainty is conducted using the Scipy 
package $curve \_ fit$ function \citep{jones01}, 
which is essentially least-squares fitting.
The result is shown in Figure~\ref{califit}.

After applying the calibration, the measurements from the calibrated four lines
(3517.299~\AA, 3545.196~\AA, 3715.464~\AA, and 3951.957~\AA)
are able to statistically match the measurements 
from the other three lines within 1$\sigma$.
However, due to the uncertainties introduced by the calibration,
these calibrated data now receive a lower weight
in determining the final abundance for each star, as shown in Figure~\ref{caliplot}.
We also point out that the same calibration 
is applied to the upper limits.

The V~\textsc{i} lines we used are in a longer wavelength
region that is not affected, so no calibrations are applied to
them.

\subsection{Final Abundances}
\label{abundances}

\begin{figure}
\begin{center}
\includegraphics[angle=0,width=3.35in]{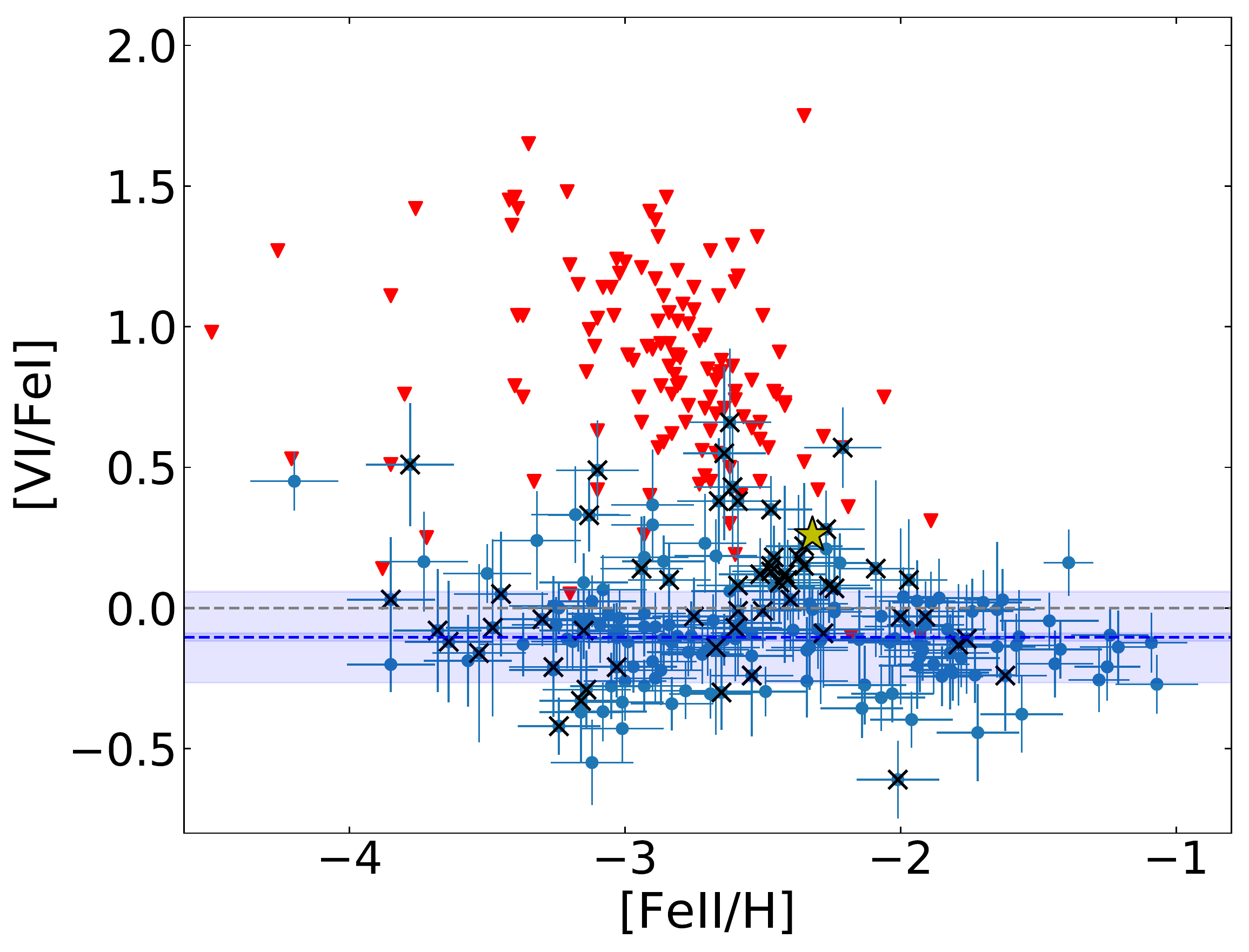} \\
\includegraphics[angle=0,width=3.35in]{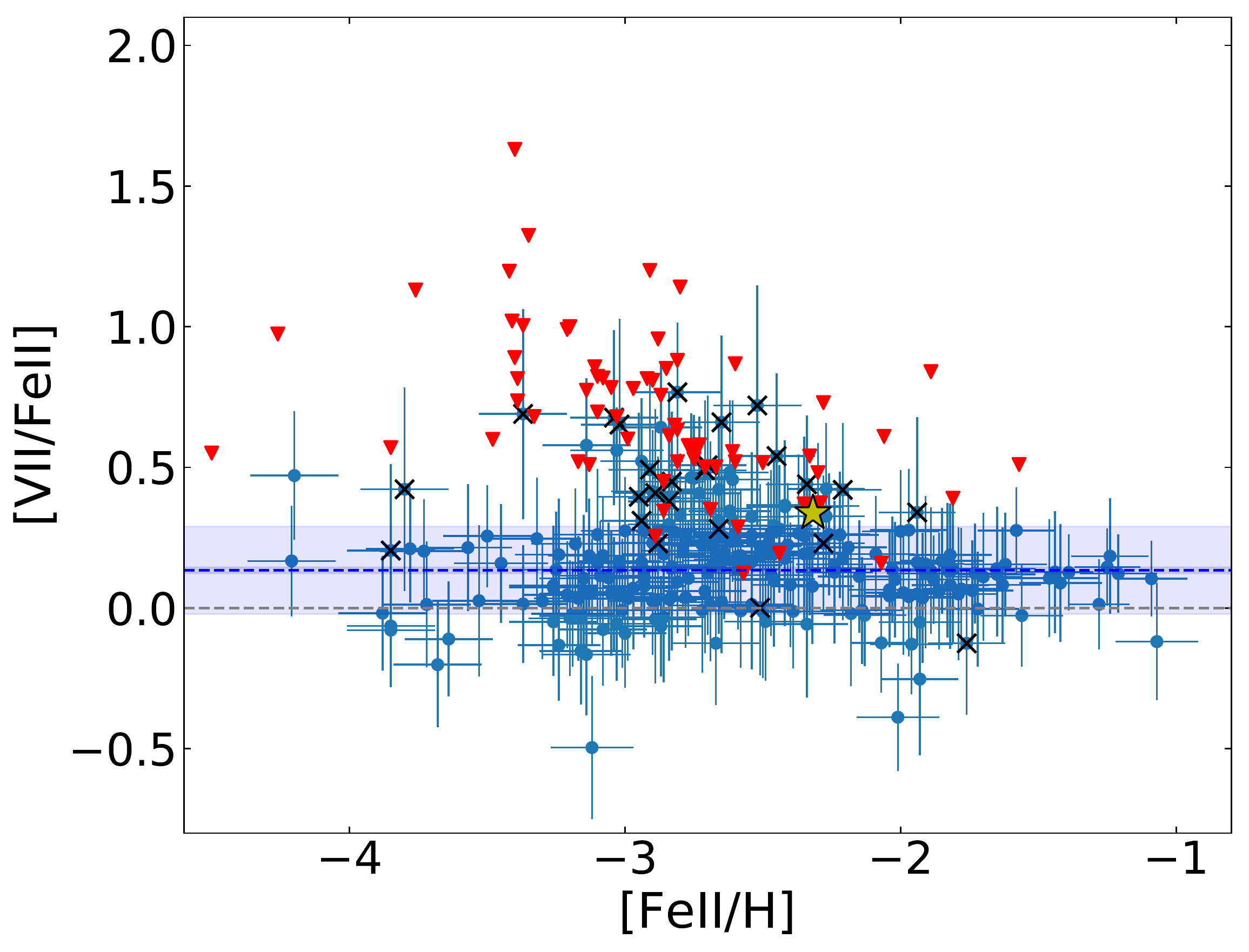}
\end{center}
\caption{
\label{metalplot}
[V/Fe] as a function of metallicity.
The blue dots are detections, 
red triangles are upper limits, 
and black crosses are stars with only one line detected.
The yellow star is HD~84937.
The blue lines represent the weighted average of the whole sample 
calculated using stars with $\geq$~2 lines of a given V species detected.
The bands represent 1$\sigma$ intervals.
 }
\end{figure}

In total, we have
190 stars in this sample with V~\textsc{i} detections
(including 10 without V~\textsc{ii}) from 4 lines,
251 stars with V~\textsc{ii} detections (including 71 without V~\textsc{i}) from 7 lines,
and 255 stars with either V~\textsc{i} or V~\textsc{ii}. 
After the calibration has been applied,
we weight the vanadium measurements by their overall
uncertainties to calculate the final abundance for each star.
We combine the measurements from the seven ionized lines,
and the four neutral lines,
to generate the ionized and neutral abundances of vanadium respectively.
The final abundances are shown in Table~\ref{vtab}.
Figure~\ref{metalplot} shows the [V/Fe] ratios as a function of metallicity.
In the metallicity range $-$4.0~$<$~[Fe/H]~$< -$1.0,
the mean [V~\textsc{i}/Fe~\textsc{i}] ratio is 
[V~\textsc{i}/Fe~\textsc{i}]~$= -0.10 \pm 0.01~(\sigma = 0.16)$~dex
from 128 stars with $\geq$~2 V~\textsc{i} lines detected,
and the mean [V~\textsc{ii}/Fe~\textsc{ii}] ratio is 
[V~\textsc{ii}/Fe~\textsc{ii}]~$= +0.13 \pm 0.01~(\sigma = 0.16)$~dex
from 220 stars with $\geq$~2 V~\textsc{ii} lines detected.

\begin{figure}
\vspace*{0.2in}  
\begin{center}
\includegraphics[angle=0,width=3.35in]{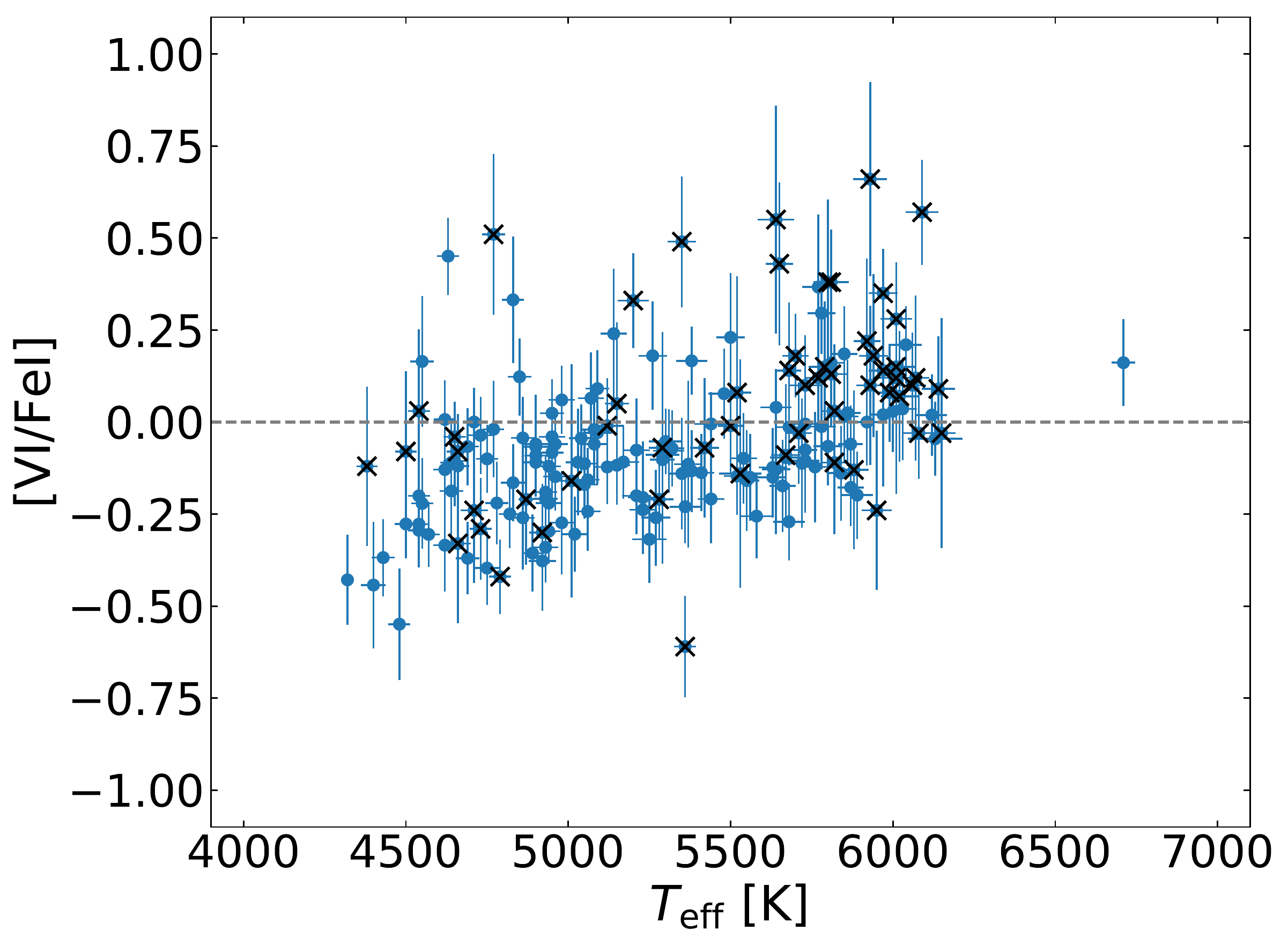} \\
\includegraphics[angle=0,width=3.35in]{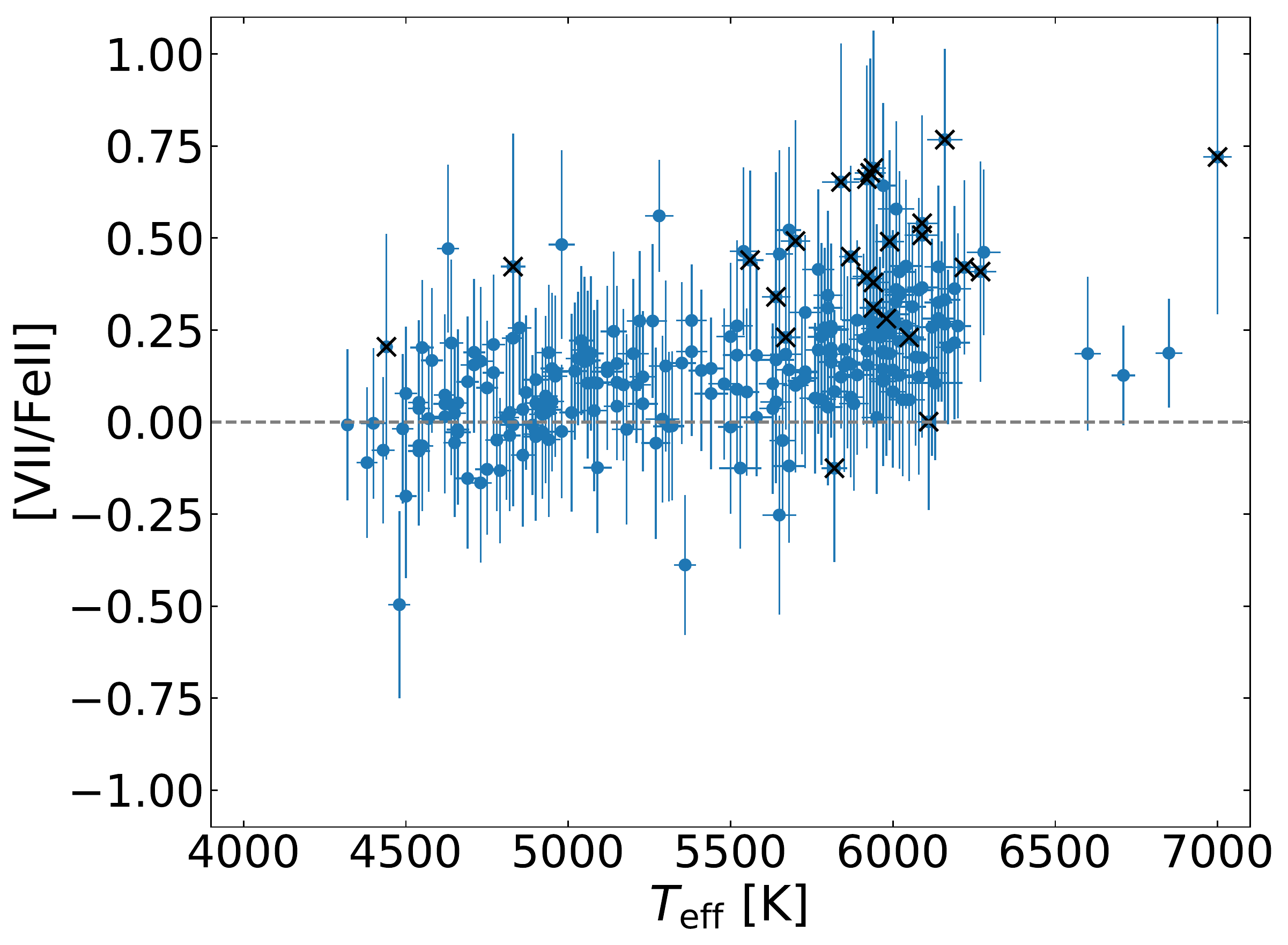}
\end{center}
\caption{
\label{teffplot}
[V/Fe] as a function of \teff.
Note that the upper limits and the stars 
with only one detected line are excluded.
 }
\end{figure}

\begin{deluxetable*}{ccccccccccccccccc}
\tablecaption{Final Abundances for the Sample
\label{vtab}}
\tablewidth{0pt}
\tabletypesize{\scriptsize}
\tablehead{
\multicolumn{6}{c}{} &
\multicolumn{5}{c}{V~\textsc{i}} &
\colhead{ } &
\multicolumn{5}{c}{V~\textsc{ii}} \\
\cline{7-11}
\cline{13-17}
\colhead{Star} &
\colhead{Class} &
\colhead{\teff (K)} &
\colhead{\logg} &
\colhead{[Fe~\textsc{i}/H]} &
\colhead{[Fe~\textsc{ii}/H]} &
\colhead{U. L.}\tablenotemark{a} &
\colhead{U. L. $\lambda$} &
\colhead{[V~\textsc{i}/Fe~\textsc{i}]} &
\colhead{$\sigma$} &
\colhead{$N$} &
\colhead{ } &
\colhead{U. L.}\tablenotemark{a} &
\colhead{U. L. $\lambda$} &
\colhead{[V~\textsc{ii}/Fe~\textsc{ii}]} &
\colhead{$\sigma$} &
\colhead{$N$}
}
\startdata
BD~$+$10\degree2495 &    RG &       4890 &       1.85 &      -2.16 &      -2.14 &          0 &         -- &      -0.36 &        0.10 &          4 &            &          0 &         -- &      -0.01 &       0.19 &          5 \\ 
BD~$+$19\degree1185a &    MS &       5440 &       4.30 &      -1.25 &      -1.25 &          0 &         -- &      -0.21 &       0.12 &          4 &            &          0 &         -- &       0.15 &       0.14 &          4 \\ 
BD~$+$24\degree1676 &    SG &       6140 &       3.75 &      -2.70 &      -2.54 &          1 &   4111.779 &       0.81 &         -- &          0 &            &          0 &         -- &       0.32 &       0.23 &          3 \\ 
BD~$+$26\degree3578 &    SG &       6060 &       3.75 &      -2.60 &      -2.41 &          0 &         -- &       0.10 &       0.14 &          1 &            &          0 &         -- &       0.22 &       0.22 &          2 \\ 
BD~$+$29\degree2356 &    RG &       4710 &       1.75 &      -1.73 &      -1.62 &          0 &         -- &      -0.24 &       0.20 &          1 &            &          0 &         -- &       0.16 &       0.18 &          4 \\ 
BD~$+$44\degree0493 &    RG &       5040 &       2.10 &      -4.28 &      -4.26 &          1 &   4111.779 &       1.27 &         -- &          0 &            &          1 &   3715.464 &       0.97 &         -- &          0 \\ 
\enddata      
\tablecomments{The complete version of Table~\ref{vtab} is available in the online edition of the journal. 
A short version is included here to demonstrate its form and content.} 
\tablenotetext{a}{Flags for upper limits. ``1'' means the [V/Fe] value given is the upper limit.}
\end{deluxetable*}

\section{Results}
\label{results}

\subsection{Trends with Stellar Parameters}
\label{trendsp}

We first examine the [V/Fe] abundance ratios 
as a function of \teff\ 
in Figure~\ref{teffplot}.
There is a difference of about 0.3~dex
in the mean
[V/Fe] ratios between
the coolest stars (around 4500~K) and
the warmest stars (around 6200~K),
with the warmest stars showing higher
[V/Fe] ratios on average.
Trends that are similar in both 
magnitude and direction are found among the 
[Sc/Fe], [Ti/Fe], [Cr/Fe], and [Mn/Fe] ratios 
in the \citet{roederer14c} sample,
leading us to conclude that the 
stellar parameters and Fe abundances 
are responsible for these trends.
Resolving the issue could require
a reanalysis of the stellar parameters
and metallicities for the entire sample,
which is beyond the scope of the present study.

\begin{figure*}
\begin{center}
\includegraphics[angle=0,width=5.00in]{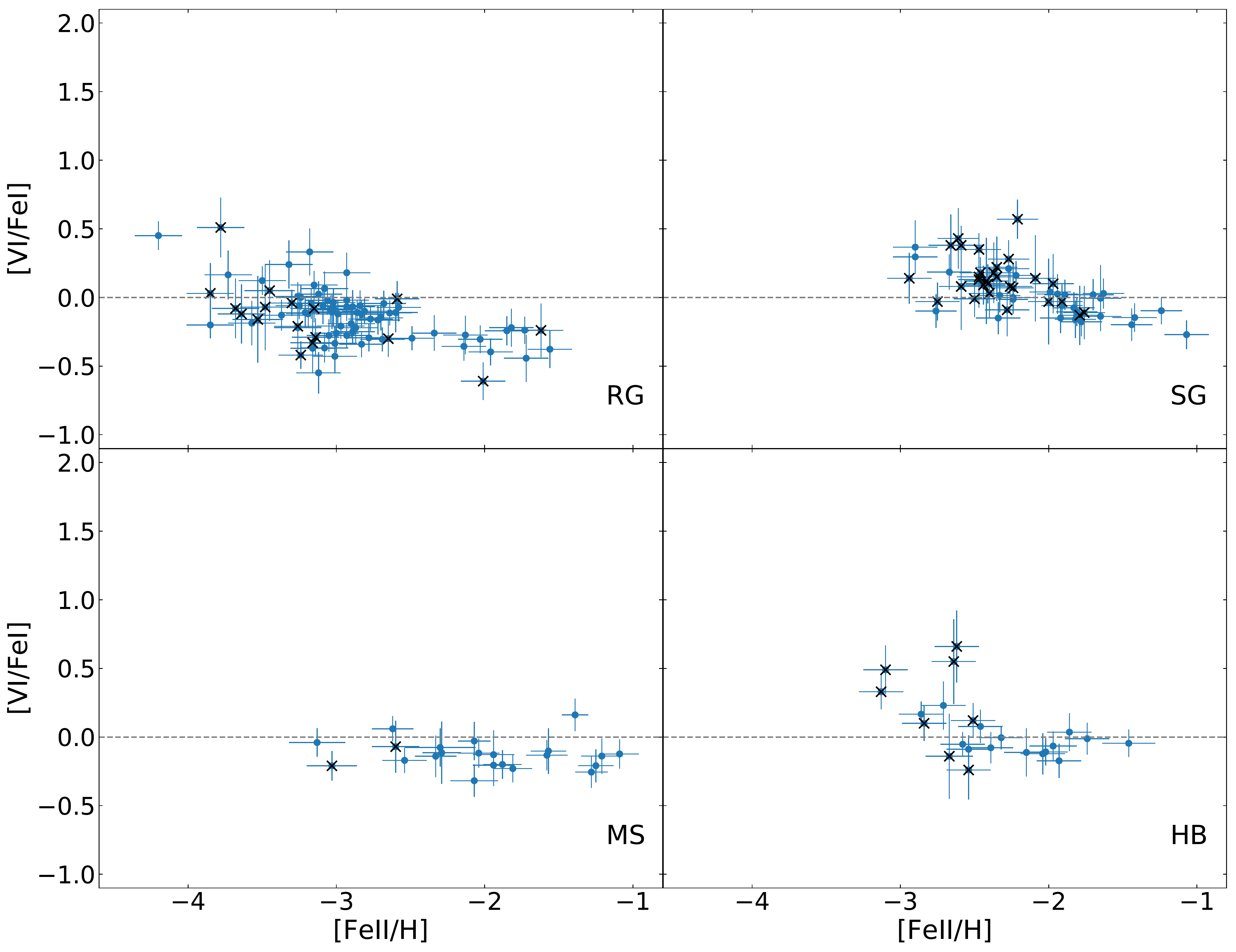} \\
\includegraphics[angle=0,width=5.00in]{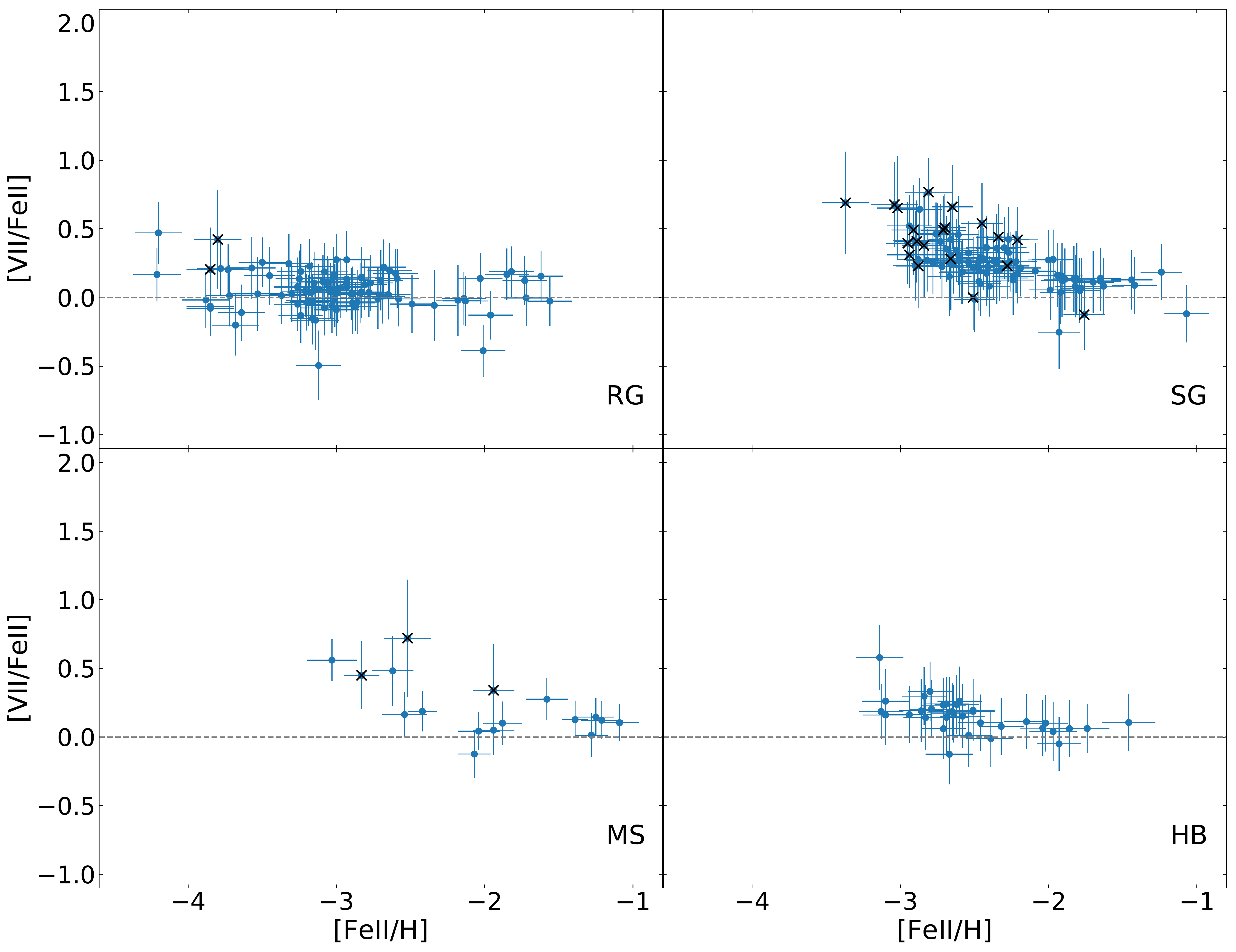}
\end{center}
\caption{
\label{metalplotclass}
[V/Fe] as a function of metallicity in separate class groups.
The legend is the same as in Figure~\ref{metalplot},
and upper limits have been omitted.
}
\end{figure*}

We also examine the [V/Fe] abundance ratios
as a function of [Fe/H] 
and stellar evolutionary state
in Figure~\ref{metalplotclass}.
We regard the [V~\textsc{ii}/Fe] ratios to be 
better representations of the [V/Fe] ratios in these stars
than the [V~\textsc{i}/Fe] ratios, which are likely 
subject to departures from LTE (Section~\ref{nonlte}).
Among the red giants in the sample, which extend
across the widest range in metallicity
(roughly $-4.0 <$~[Fe/H]~$< -1.5$), 
there is no dependence of [V~\textsc{ii}/Fe] on [Fe/H].

The trend of increasing [V/Fe] ratios with decreasing metallicity among subgiants
is likely caused by the fact that we need a much stronger line to 
get a valid detection from
these warm, low-metallicity stars.
The higher the temperature, 
as well as the lower the metallicity and vanadium abundance,
the higher that limit is.
We perform a separate test to determine where those limits lie.
We generate model atmospheres 
at several typical temperatures and metallicities 
that resemble the actual sample.
Using these hypothetical model atmospheres,
we determine the minimum detectable abundance 
based on our synthetic spectra.~
The stars with one or no vanadium lines detected
are consistent with these tests.

Many of the subgiants at the lower metalliciy end
in Figure~\ref{metalplotclass}
have only one detection from the strongest line.
Since these stars have only one detection,
and that detection lies near the limit for 
minimum detectable vanadium abundances,
we decide to exclude them from subsequent analyses.
When we examine correlations between vanadium 
and other iron-group elements (Section~\ref{correlation}), 
these stars are
not included or shown in the figures.

We also investigate the impact of the stellar parameter
uncertainties on the
derived [V/Fe] ratios.
For this test,
we select two representative red giants with similar 
stellar parameters and metallicities 
(\object[HD 21581]{HD~21581}:\ 
\teff\ = 4940~K, \logg\ = 2.10, \vt\ = 1.50~\kmsec, [Fe/H] = $-$1.82;
\object[HD 126238]{HD~126238}:\ 
\teff\ = 4750~K, \logg\ = 1.65, \vt\ = 1.55~\kmsec, [Fe/H] = $-$1.96) and
contrasting vanadium abundances
([V~\textsc{ii}/Fe~\textsc{ii}] = $+$0.19, 
derived from 6 V~\textsc{ii} lines; and
[V~\textsc{ii}/Fe~\textsc{ii}] = $-$0.13, 
derived from 7 V~\textsc{ii} lines, respectively).
We resample the stellar parameters $10^{3}$ times
assuming normal error distributions
on \teff, \logg, \vt, and [M/H],
and we recompute the vanadium abundances for each new combination of parameters.
The resamples include the dependence of \logg\ on \teff\ through
the $\text{Y}^{2}$ isochrones.
For this test, we approximate the abundances derived 
via spectrum synthesis as equivalent widths.

\begin{figure}
\begin{center}
\includegraphics[angle=0,width=3.35in]{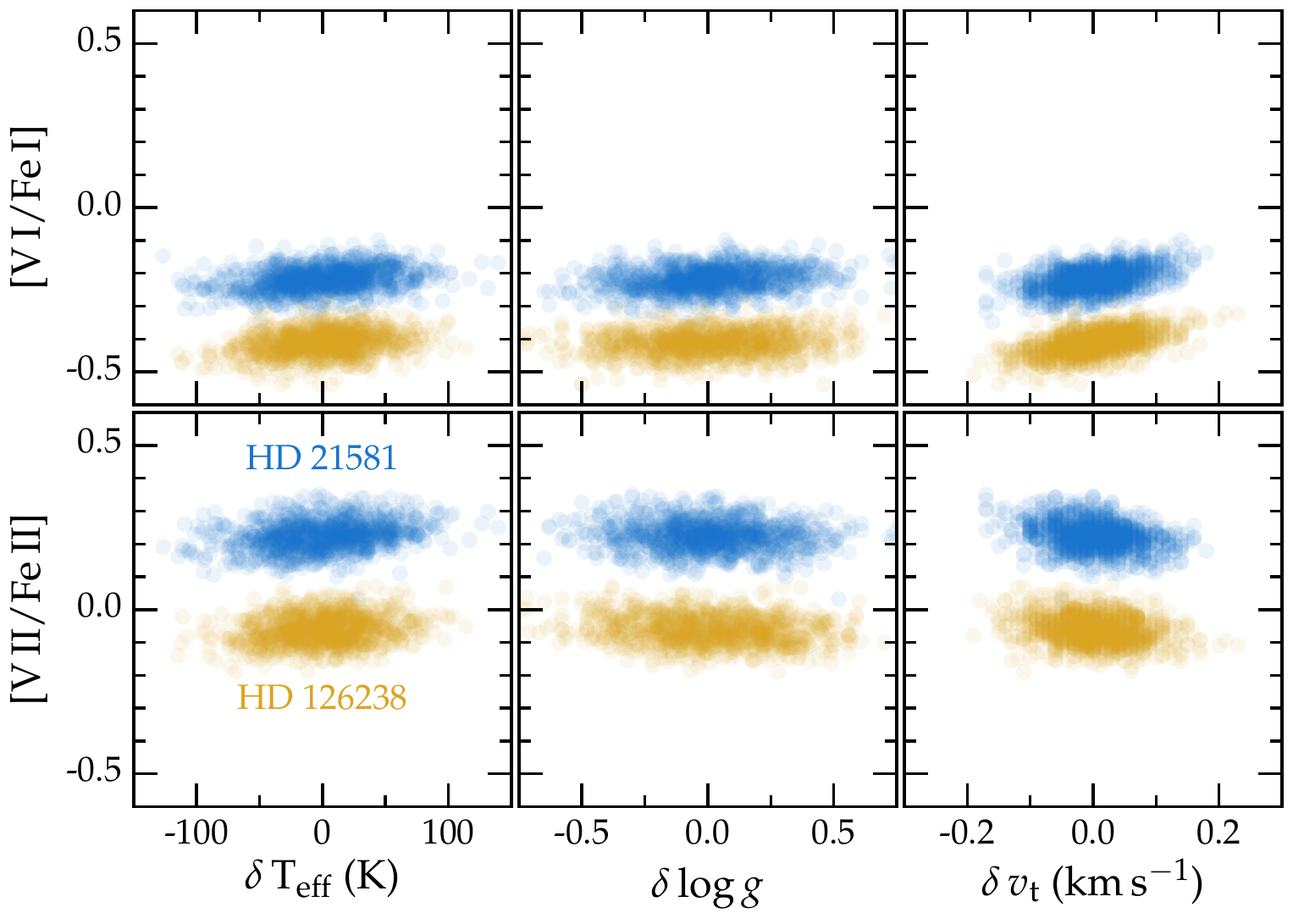} 
\end{center}
\caption{
\label{stellarparvarplot}
Comparison of the [V/Fe] ratios for two representative red giants
as a function of changes in the stellar parameters.
The blue points represent the results for
\object[HD 21581]{HD~21581}, and 
the yellow points represent the results for
\object[HD 126238]{HD~126238}.
The three columns represent changes in \teff, \logg, and \vt\
relative to the median value of each parameter.
Notice that there is minimal change in the [V/Fe] ratio for each
star, whether derived from lines of the neutral or ionized species,
and the high value of [V/Fe] in 
\object[HD 21581]{HD~21581} robustly remains in contrast to
the low value of [V/Fe] in 
\object[HD 126238]{HD~126238}.
}
\end{figure}

Figure~\ref{stellarparvarplot} illustrates the results of this test.
Three important features are found.
First, the [V/Fe] ratios from lines of both
neutral and ionized species show
minimal sensitivity to variations in the stellar parameters.
Secondly, the [V/Fe] ratio is consistently
higher when derived from
lines of ionized atoms than lines of neutral atoms
(see also Section~\ref{nonlte}).
Finally, the contrast between the high [V/Fe] ratio in 
\object[HD 21581]{HD~21581} 
and the low [V/Fe] ratio in 
\object[HD 126238]{HD~126238}
is always present, suggesting that there are 
intrinsic [V/Fe] differences from one star to another
(see also Section~\ref{correlation}).
These three behaviors are largely unaffected by
uncertainties in the stellar parameters.

\subsection{Possible Non-LTE Effects}
\label{nonlte}

\begin{figure}
\begin{center}
\includegraphics[angle=0,width=3.35in]{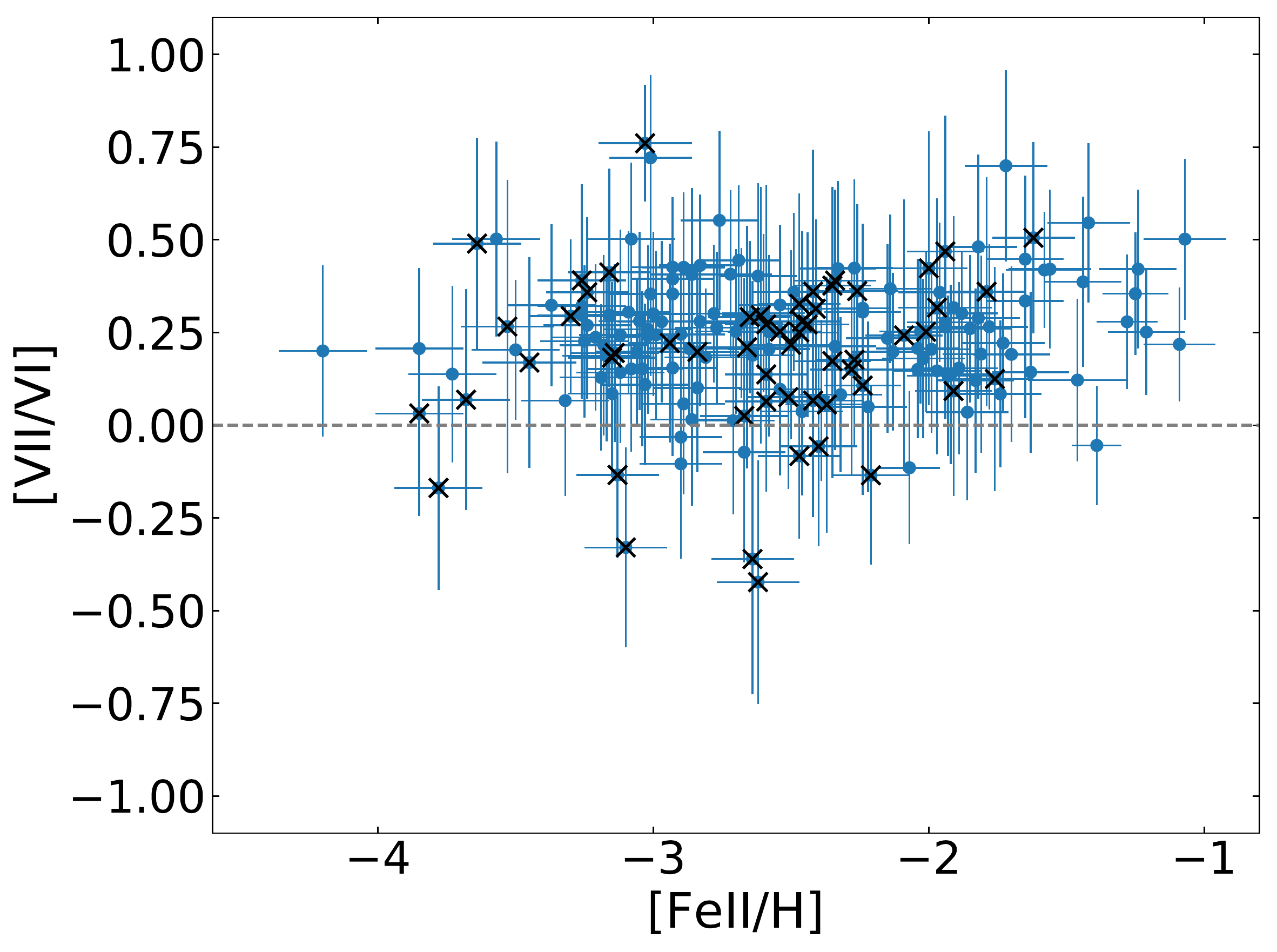} \\
\includegraphics[angle=0,width=3.35in]{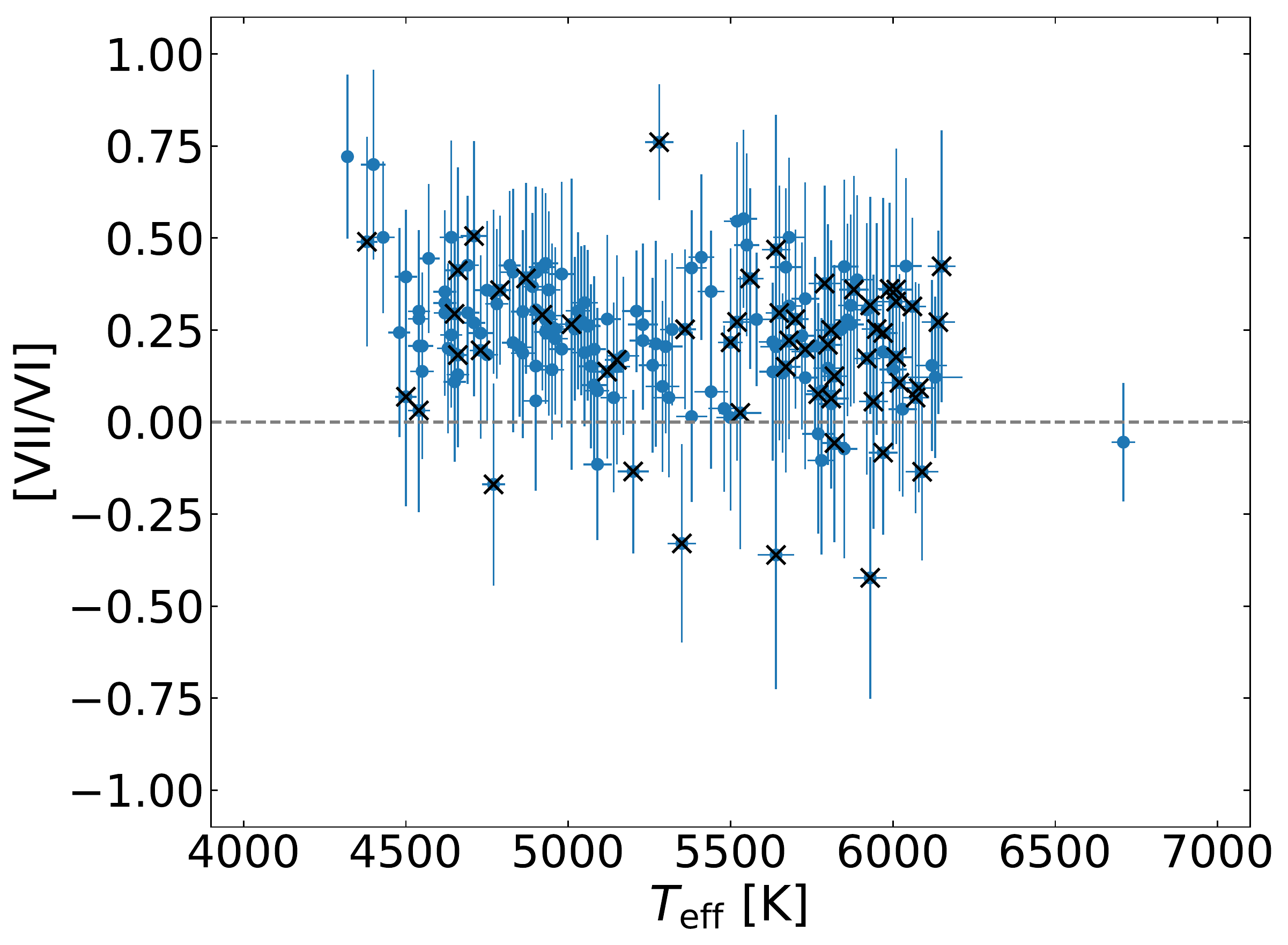}
\end{center}
\caption{
\label{v2v1plotclass}
Comparison of [V~\textsc{ii}/V~\textsc{i}] as
a function of [Fe/H] (top) and \teff\ (bottom).
 }
\end{figure}

We examine possible non-LTE effects by comparing the
vanadium abundance derived from neutral vanadium and that
from ionized vanadium.
Our calculations are made assuming LTE Saha ionization equilibrium.
We see in Figure~\ref{v2v1plotclass} that the neutral vanadium lines
give slightly lower
vanadium abundances than the ionized vanadium lines,
[V~\textsc{ii}/V~\textsc{i}]~$= +0.25 \pm 0.01~(\sigma = 0.15)$~dex
from 119 stars.
This suggests that non-LTE effects may be important, and
we assume that over-ionization of neutral vanadium is
occurring.
No non-LTE studies of vanadium abundances in late-type stars
have been conducted,
and we encourage non-LTE calculations for neutral vanadium
in the future 
to test this hypothesis and obtain 
more accurate results for abundances derived from V~\textsc{i} lines.

The V~\textsc{ii} lines we use are connected to the low 
metastable levels of the ion,
which are the primary population reservoir levels for vanadium in the atmospheres of late-type stars.
Therefore, we consider the abundances derived from the V~\textsc{ii} lines
to be less affected by non-LTE effects, and they should provide
more accurate abundances
(cf.\ \citealt{lawler13}).

When calculating the vanadium abundance ratio using the neutral lines,
we use the iron abundance derived from the neutral iron species
for consistency, 
and vice versa for the abundance ratio derived from the ionized vanadium lines.
We note this fact by writing [V~\textsc{i}/Fe~\textsc{i}] and 
[V~\textsc{ii}/Fe~\textsc{ii}] when possible.
Furthermore, when we examine the correlations between the 
vanadium abundances derived in this study and other iron-group elements
from \citet{roederer14c} (Section~\ref{correlation}),
we compare abundances derived from neutral iron-group species with those from neutral vanadium
and abundances derived from ionized iron-group species with those from ionized vanadium.

\subsection{Correlation with Other Iron-group Elements}
\label{correlation}

We observed in Section~\ref{trendsp} 
that there are intrinsic differences in the [V/Fe] ratios
found among this sample of stars, and
we showed that these differences are not
a consequence of uncertainties in the stellar parameters.
These differences also cannot be attributed to the \loggf\ values.
Systematic uncertainties in the laboratory \loggf\ values 
for all of the lines of iron-group elements considered here---including
scandium, titanium, and vanadium---are small
(typically $<$~5~\% or $<$~0.02~dex), so
they are effectively negligible in the abundance error budget.

Following \citet{sneden16} and \citet{cowan20}, 
we examine whether these differences may be 
related to correlations 
among scandium, titanium, and vanadium.
For this test, we adopt the abundances
of other iron-group elements presented by \citet{roederer14c}.
For completeness, we examine whether correlations exist between all other pairs of iron-group elements.
We check for correlations in the [X/Fe] ratios, where X represents
Sc~\textsc{ii}, Ti~\textsc{i}, Ti~\textsc{ii}, V~\textsc{i}, V~\textsc{ii}, 
Cr~\textsc{i}, Cr~\textsc{ii}, Mn~\textsc{i}, Mn~\textsc{ii},
Co~\textsc{i}, and Ni~\textsc{i}.

\begin{figure}
\begin{center}
\includegraphics[angle=0,width=3.35in]{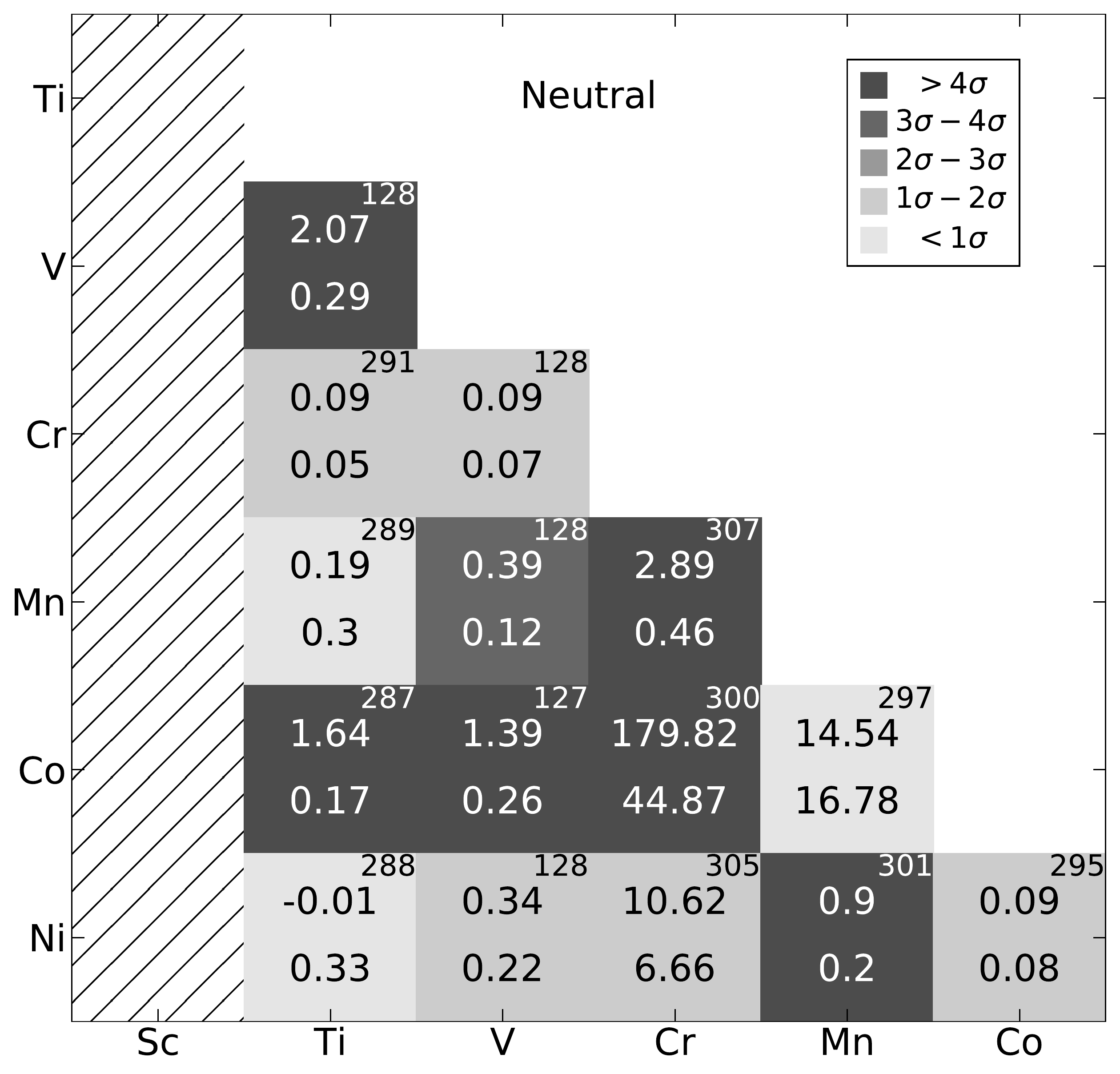} \\
\includegraphics[angle=0,width=3.35in]{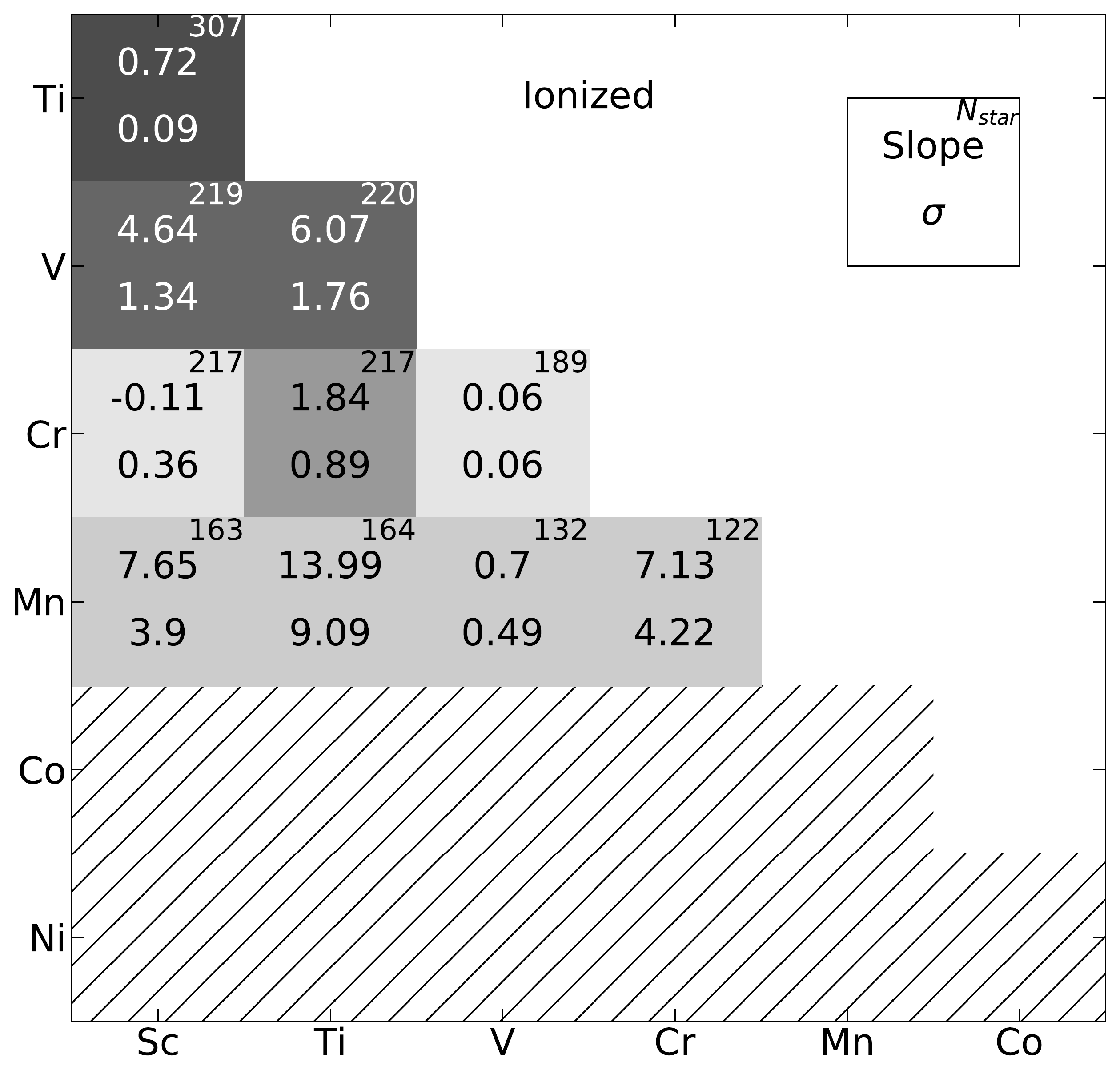}
\end{center}
\caption{
\label{main_slope}
A summary of the slopes for the correlation fittings between
11 iron-group species, expressed as [X/Fe], with neutral species in the upper panel and
ionized species in the lower panel.
The color of the squares represent the significance of the correlation,
as indicated in the upper panel.
The meaning of the numbers are given in the lower panel,
where slope is the median of the slopes from 1000 resampling,
$\sigma$ is distance from the median to the 16~\%\ and 84~\%\ intervals,
and $N_{star}$ is the number of stars used in each correlation 
calculation.
The hatching indicates correlations where we have no available abundances
to perform the fitting.
}
\end{figure}

\begin{figure*}
\begin{center}
\includegraphics[angle=0,width=6.8in]{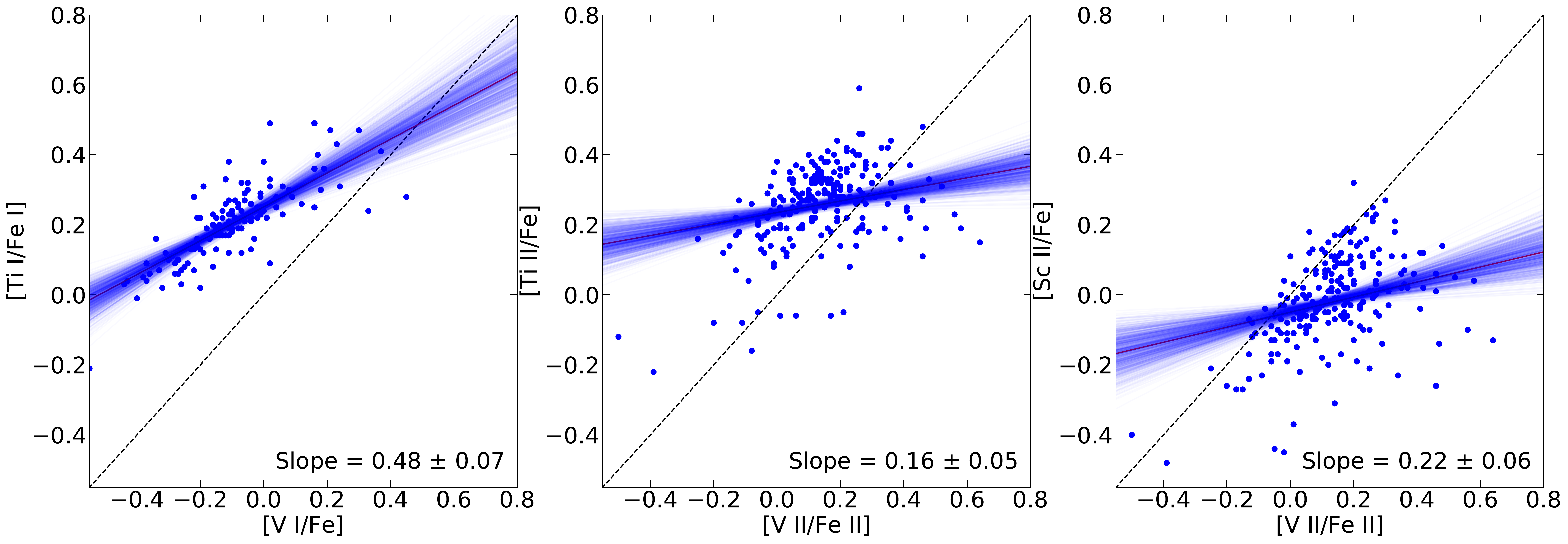}
\end{center}
\caption{
\label{sctiv_slope}
Correlation fitting for three pairs of iron-group elements.
The panels show the high-significance correlations between
scandium, titanium, and vanadium.
The best fit recorded in Figure \ref{main_slope} is plotted as the
solid line, with all 1000 re-samplings plotted in low opacity.
A 45\degree\ dashed line is included for comparison.
Note the axes are swapped as compared to the correlation
fits shown in Figure~\ref{main_slope} to keep vanadium consistently
on the x-axis.
The slope printed at the lower right corner of each panel is changed
accordingly, but they maintain the same significance as a result of
ODR.
}
\end{figure*}

Since the abundances are all derived quantities,
we apply an orthogonal distance regression (ODR) 
to the sample to obtain the 
slope as a measure of the correlation.
To incorporate the uncertainties in the measurements,
we generate 1000 re-samplings of the measurements 
based on the uncertainties, and we
apply ODR to the re-sampled data.
Because the slope distributions for most pairs
of elements are not Gaussian,
the medians are taken as the final results, with
the 16~\%\ and 84~\%\ intervals as the standard deviations.
We compute the significance of the correlation as the slope
divided by the standard deviation.
Other fitting methods, such as the merit function maximization and 
the Markov chain Monte Carlo (MCMC) method, yield consistent results.
The fitting results over the entire sample are shown in
Figure~\ref{main_slope}.
The figure shows the significance of the correlation between each
possible pair of iron-group elements by the shade.
For example, the slope of [V~\textsc{i}/Fe] versus [Ti~\textsc{i}/Fe]
is $2.07 \pm 0.29$ with a significance greater than 7$\sigma$, 
whereas [Cr~\textsc{ii}/Fe] versus [V~\textsc{ii}/Fe]
has slope $0.06 \pm 0.06$ with a significance less than 1$\sigma$.
Among all the correlations involving vanadium,
we see more significant correlations with
ionized scandium and 
neutral as well as ionized titanium.
The ODR gives significance for these correlation greater than 3$\sigma$.
We present the three correlation fits between scandium, titanium,
and vanadium
in Figure~\ref{sctiv_slope}
to better illustrate the significance of the correlation
slope.
We also apply the ODR to subgroups divided according to metallicities
and evolutionary stages and obtain results that agree with
the behavior of the full sample.

These correlations among 
scandium, titanium, and vanadium
cannot be a consequence of the correlations with \teff\ that were noted in Section~\ref{trendsp}.
If they were, then we would also expect to see correlations between these elements and chromium
and manganese; 
no significant correlations are detected between 
chromium or manganese and
scandium or titanium, and there are other reasons to disfavor
those between vanadium, chromium, and manganese (see below).
Furthermore, as shown by \citet{cowan20}, 
correlations among 
scandium, titanium, and vanadium 
are detected in multiple independent datasets
whose stars do not show any correlations with \teff.

\begin{figure*}
\begin{center}
\includegraphics[angle=0,width=6.8in]{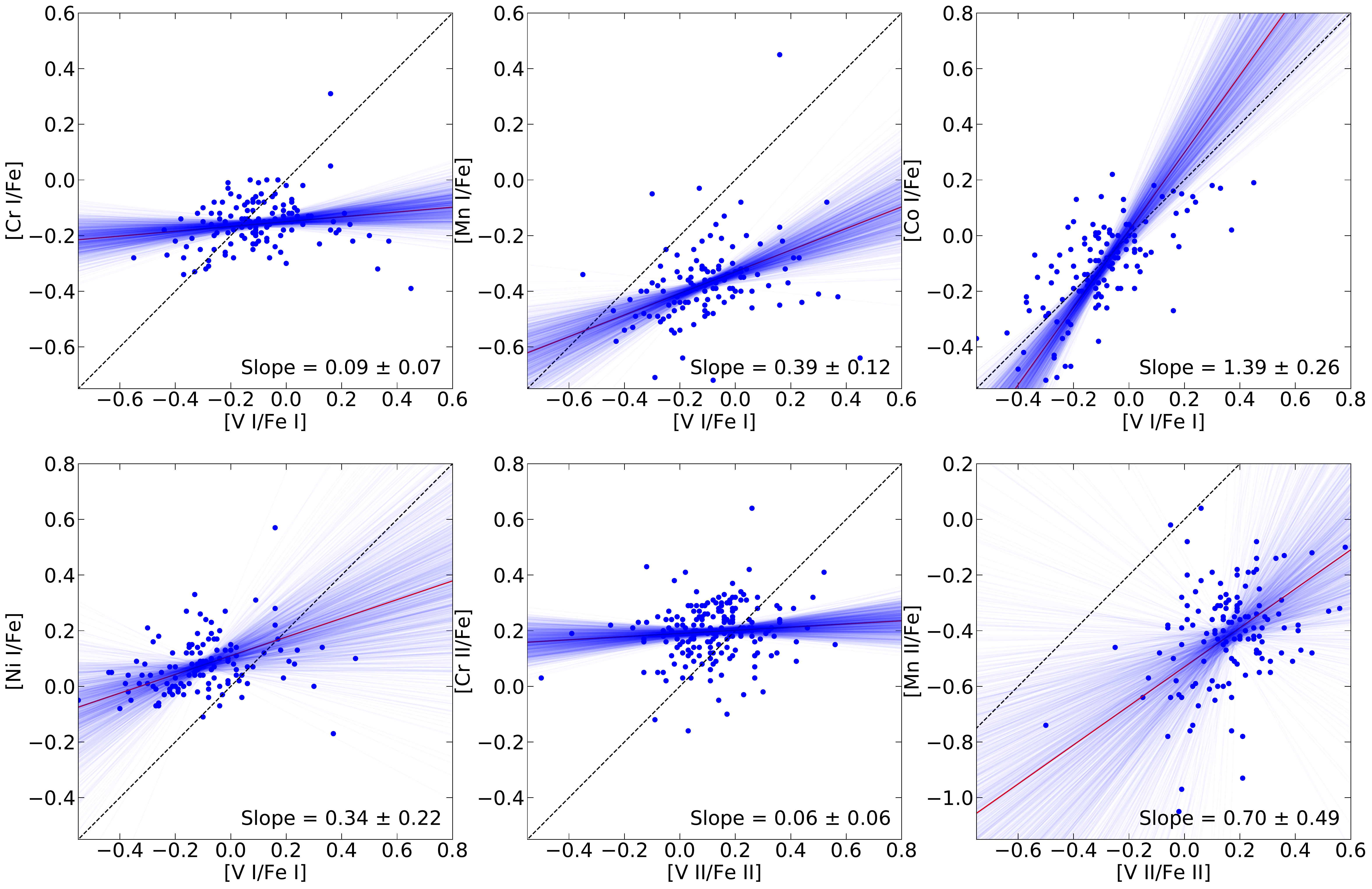}
\end{center}
\caption{
\label{vcrmnconi_slope}
Correlation fitting for six pairs of iron-group elements.
The panels show the correlations between
vanadium, chromium, manganese, cobalt, and nickel.
The first, fourth, fifth, and sixth correlations are
not significant,
whereas the second and third are significant.
For reasons stated in the text, we conclude the correlations
among vanadium, manganese, and cobalt are not reliable.
}
\end{figure*}

We also notice significant correlations between cobalt, manganese, 
and other elements.
In particular, [Co~\textsc{i}/Fe~\textsc{i}] shows significant correlations
with [Ti~\textsc{i}/Fe~\textsc{i}], [V~\textsc{i}/Fe~\textsc{i}], 
and [Cr~\textsc{i}/Fe~\textsc{i}].
We present the correlation fits of vanadium with chromium,
manganese, cobalt, and nickel in Figure~\ref{vcrmnconi_slope}.
There are, however, reasons to discount these other correlations.
As pointed out by \citet{cowan20}, neutral cobalt, which is a trace species,
yields substantially higher abundances
than ionized cobalt in the three metal-poor stars examined in that study.
Non-LTE effects on the abundance derived from
Co~\textsc{i} lines can be substantial
\citep{bergemann10}.
\citeauthor{cowan20}\ concluded that the abundances derived from neutral
cobalt likely are not correct.
Although we do not have cobalt measurements 
from Co~\textsc{ii} lines to make the
same comparison for our sample,
it is likely that our neutral cobalt measurements are affected by the same issue.
Any correlations between cobalt and other iron-group elements found here
should be treated with caution.

Non-LTE effects may also be responsible for 
correlations involving neutral titanium, vanadium, chromium, 
manganese, and nickel.
These elements have low first ionization potentials
($<$ 7.7~eV), so their 
neutral species are distinct minorities
in the atmospheres of the stars in our sample.
Previous calculations have shown
that non-LTE effects can be significant
(e.g., 
Ti:\ \citealt{sitnova20};
Cr:\ \citealt{bergemann10cr};
Mn:\ \citealt{bergemann08}).
No non-LTE studies of nickel abundances are available, 
but nickel has a similarly low first ionization potential,
so levels of its neutral species could reasonably
be expected to deviate from their LTE populations.

These concerns motivate our recommendation
to favor the correlation results derived from
the majority ion species and disfavor the results derived from
the neutral species.
As shown in the bottom panel of Figure~\ref{main_slope},
scandium, titanium, and vanadium correlate significantly with each other,
and similar correlations are not found among chromium or manganese.
We conclude these observed correlations most likely
arise from the production mechanism rather than systematic or
observational effects.


%

\section{Discussion}
\label{discussion}

The super-solar [V~\textsc{ii}/Fe] ratio
indicates a potential dominance of massive Type~II SNe in 
early Galactic nucleosynthesis, because vanadium is more abundantly 
produced by more massive Type~II SNe \citep{Chieffi04}.
Environments with more massive Type~II SNe would also explain 
the super-solar ratios of titanium.  
The correlated scandium, titanium, and vanadium abundances 
suggest that these three elements may be produced under similar conditions. 
Titanium is sometimes referred to as an $\alpha$ element.  
It seems unlikely, however, that these three elements are produced 
along with the $\alpha$ elements in the SNe that 
enriched the stars in our sample.  
As discussed in \citet{curtis19}, \citet{cowan20}, and references therein, 
the nucleosynthesis origin of these elements in Type~II SNe 
occurs mainly in proton-rich inner layers during 
complete silicon burning.  
Their production is complex, 
and current models remain unable to fully reproduce 
the observed abundance ratios.

The correlations we identify are not related to those 
found among more metal-rich stars.
In near-solar metallicity stars, for example,
\citet{hasselquist17}
pointed out that
vanadium and manganese are expected to be more
metallicity-dependent than iron, causing the relative abundances
of these two elements to increase with metallicity \citep{andrews17}.
It is indeed observed that [V/Fe] and [Mn/Fe] increase with [Fe/H]
in the Milky Way (MW) and Sagittarius (Sgr) dwarf galaxy samples, 
although both ratios are deficient in Sgr relative to the MW.~
This metallicity-dependent correlation is different from the
one identified in the present study,
which focuses on much more metal-poor stars 
where minimal metallicity dependence is 
observed in the [V/Fe] ratios.


\section{Conclusions}
\label{conclusion}

Using improved atomic transition data,
we rederived the vanadium abundances for 255 metal-poor stars 
from \citet{roederer14c}.
With an increase from 3 to 11 lines used, we obtained consistent measurements
with the previous study but smaller uncertainties.
As shown in Section~\ref{results}, 
our work further supports the existence of 
correlations between scandium, titanium, and vanadium in metal-poor stars.
With many more lines used, we have greater confidence in the 
reliability of these rederived vanadium abundances.
By examining trends of individual lines and overall abundances
with stellar parameters,
we have identified no systematic effects in the analysis 
that could account for the correlations observed.
We conclude that
the correlations are, therefore, very likely a result of nucleosynthesis 
at early Galactic time.

As summarized in Section~\ref{intro},
we do not yet have an explanation
for the correlations, especially between scandium, titanium, and vanadium.
The complex production history of scandium in CCSNe makes it difficult to 
understand how it is related to vanadium and titanium \citep{sneden16}.
It is possible to reproduce the overall iron-group
element abundances
by including a larger hypernova fraction and taking into account
the effect of neutrino interactions and jets
\citep{umeda02,kobayashi06,kobayashi11,curtis19}.
Yet, it is still unclear how the current CCSNe models will impact the 
GCE calculations, and whether they can reproduce the correlations seen
in observations.
More theoretical works are needed to explain the correlations
between scandium, titanium, and vanadium.

\acknowledgements

We thank G.\ Preston for obtaining some of the spectra used in this work
and the referee for offering helpful suggestions.
I.U.R.\ acknowledges funding from the US National Science Foundation (NSF)
through grants AST~1613536 and AST~1815403.
C.S.\ acknowledges funding from NSF through grant AST~1616040.
J.E.L.\ acknowledges funding from NASA through grant NNX16AE96G and from NSF
through grant AST~1814512.
This research has made use of NASA's
Astrophysics Data System Bibliographic Services;
the arXiv pre-print server operated by Cornell University;
the SIMBAD and VizieR
databases hosted by the
Strasbourg Astronomical Data Center;
and the Atomic Spectral Database hosted by the National Institute of Standards and Technology \citep{kramida18}.

\facility{%
HET (HRS), 
Magellan:Baade (MIKE), 
Magellan:Clay (MIKE), 
Smith (Tull)}

\software{%
matplotlib \citep{hunter07},
MOOG \citep{sneden73},
numpy \citep{vanderwalt11},
scipy \citep{jones01}}

\clearpage

\bibliographystyle{aasjournal}
\bibliography{xou}

\end{document}